\documentclass[aps,prb,reprint,longbibliography,superscriptaddress]{revtex4-2}
\usepackage{mathrsfs}
\usepackage{amsmath,gensymb}
\usepackage{amsfonts}
\usepackage{amssymb}
\usepackage{amsthm}
\usepackage{graphicx}
\usepackage{natbib}
\usepackage{color}
\usepackage{hyperref}
\usepackage{bm}
\usepackage[caption=false]{subfig}
\usepackage{verbatim}
\usepackage[normalem]{ulem}
\usepackage{soul}

\begin{document}
\title{Magnetic proximity effect in superconductor/ferromagnet van der Waals heterostructures: dependence on the number of superconducting monolayers}

\author{A. S. Ianovskaia}
\affiliation{Moscow Institute of Physics and Technology, Dolgoprudny, 141700 Moscow region, Russia}

\author{G. A. Bobkov}
\affiliation{Moscow Institute of Physics and Technology, Dolgoprudny, 141700 Moscow region, Russia}

\author{A. M. Bobkov}
\affiliation{Moscow Institute of Physics and Technology, Dolgoprudny, 141700 Moscow region, Russia}

\author{I.V. Bobkova}
\affiliation{Moscow Institute of Physics and Technology, Dolgoprudny, 141700 Moscow region, Russia}
\affiliation{National Research University Higher School of Economics, 101000 Moscow, Russia}

\begin{abstract}
The magnetic proximity effect in superconductor/ferromagnet (S/F) heterostructures with a large number of atomic layers leads to a suppression of the superconducting order parameter and appearance of Zeeman-like spin splitting of the local density of states (LDOS). Here we study the magnetic proximity effects in van der Waals S/F heterostructures with a few  atomic layers and demonstrate that the corresponding physics is very  different from the classical results. We find that the dependence of the superconducting order parameter exhibits dips as a function of the ferromagnetic exchange field and gating. The number of dips is determined by the number of monolayers in the heterostructure and, in general, the superconductivity is not suppressed by large values of the exchange field. The spin splitting of the LDOS cannot be described by an effective Zeeman field and manifests a multiple peak structure, where each peak is connected to a unique spin splitting of one of the superconducting bands, which also can be tuned by gating. 
\end{abstract}

\maketitle

\section{Introduction}
It is well established that in superconductor/ferromagnet (S/F) heterostructures the superconducting state is strongly modified near the S/F interface due to the magnetic proximity effect. The first essential modification is a partial conversion of parent singlet superconducting correlations to triplet ones \cite{Buzdin2005,Bergeret2005}. The second key feature is the Zeeman splitting of the  electronic local density of states (LDOS)\cite{Bergeret2018review,Heikkila2019review}, which allows observation of a number of promising phenomena in the field of superconducting spintronics \cite{Linder2015,Eschrig2015}, caloritronics and spin caloritronics\cite{Bergeret2018review,Heikkila2019review}. Some striking examples include 
giant thermoelectric \cite{Machon2013,Ozaeta2014,Kolenda2016,Kolenda2016_2,Giazotto2014,Giazotto2015,Machon2014,Kolenda2017,Rezaei2018}, thermospin effects
\cite{Ozaeta2014,Linder2016,Bobkova2017}, highly efficient domain wall motion induced via temperature gradients \cite{Bobkova2021,Bobkov2021}, cooling
at the nanoscale \cite{Giazotto2006review,Kawabata2013}, spin and heat valves \cite{Huertas-Hernando2002,Giazotto2006,Giazotto2008,Giazotto2013,Strambini2017},  low-temperature thermometry \cite{Giazotto2015_2} and
detectors of electromagnetic radiation \cite{Heikkila2018,Geng2023}.

The magnetic proximity effect is well studied in heterostructures, with a large number of atomic layers, such that the
properties of the S and F layers are close to the properties of the corresponding bulk materials and effects of an interface hybridization of electronic bands of the layers are not important. However, the discovery of 2D materials that opens up unprecedented opportunities for the design of new functional materials requires the development of a theory of the magnetic proximity effect in van der Waals (vdW) S/F heterostructures. This became the motivation for this work. 

At first, as a background, we will briefly formulate the main features of the classical magnetic proximity effect in thin-film heterostructures, with a large number of atomic layers. Here the term "thin-film" means that the thickness of the S layer is still much smaller than the superconducting coherence length, $d_S \ll \xi_S$. In this case the magnetic proximity to  ferromagnetic metals can be described by an  effective depairing parameter accounting for a possible leakage of the superconducting correlations into the F layer and subsequent destruction there and by an effective exchange field induced in the S layer.  The effective exchange field also suppresses superconductivity and leads to the Zeeman splitting of the LDOS in the superconductor.  For example, if the S layer is proximitized by a thin-film metallic ferromagnet and the S/F interface is fully transparent, the effective exchange field in the S layer is $h_{eff} = \nu_F d_F h/(\nu_S d_S + \nu_F d_F)$, and the leakage of the Cooper pairs is described by a reduced effective coupling constant $\lambda_{eff} = \nu_S d_S \lambda/(\nu_S d_S + \nu_F d_F)$. Here $\lambda$ is the superconducting coupling constant of the isolated S layer and $h$ is the exchange field of the isolated ferromagnet, $\nu_{S(F)}$ is the density of states at the Fermi level in the S (F) layer \cite{deGennes1964,Bergeret2001}.  

In the opposite case of a superconductor proximitized by a ferromagnetic insulator the magnetic proximity effect is described by the so-called spin-mixing angle \cite{Tokuyasu1988,Cottet2009}, which is produced by spin-dependent electron scattering from the S/F interface and is determined by the exchange field of the ferromagnetic insulator. Expansion of the boundary conditions to the linear
order in the spin-mixing angle   is equivalent to appearance of the effective exchange field in the  thin S layer, and the
second order is equivalent to pair breaking by spin-dependent scattering \cite{Cottet2009,Eschrig2015_bc}.  Analogously to the case of proximity with a ferromagnetic metal the effective exchange field leads to the Zeeman splitting of the LDOS and both mentioned factors suppress superconductivity. Both in the case of a ferromagnetic metal and in the case of a ferromagnetic insulator  it is expected that a stronger ferromagnet produces a stronger magnetic proximity effect in the S layer and, therefore, stronger suppression of superconductivity. 

Although different aspects of proximity effects in S/F vdW heterostructures are being studied very actively \cite{Aikebaier2022,Kang2021,Wickramaratne2021,Jo2023,Jiang2020,Kezilebieke2020,Ai2021,Idzuchi2021}, the influence of proximity to a few- or monolayer ferromagnet on the superconductivity in a few-layer vdW superconductor has not yet been studied. Recently the magnetic proximity effect was studied in S/F bilayer systems consisting of monolayer van der Waals (vdW) materials \cite{Bobkov2024_vdW}. It was found that in this case the magnetic proximity effect is determined by the hybridization of the electronic bands of both materials. The important consequence of this mechanism is that both the effective exchange field and the depairing parameter due to the leakage of the Cooper pairs into the F layer can be controlled via gating. In particular, the amplitude and the sign of the effective exchange field can be changed by applying the gating potential,  which looks very promising from the point of view of the potential use of such structures in spintronics and spin caloritronics. For example, a possibility to generate a gate controllable superconducting spin current in such structures was reported in \cite{Bobkov2024_spin}.

In this paper we study the dependence of the magnetic proximity effect in vdW S/F heterostructures on the number of superconducting layers $N$. We demonstrate that the hybridization mechanism of the magnetic proximity effect clearly manifests itself with change of the number of layers. It is shown that heterostructures with 2-3 superconducting layers have the most pronounced non-monotonic dependence of the superconducting order parameter on the exchange field of the F layer. This non-monotonic dependence reveals several dips, whose number is determined by the number of the S layers. Moreover, it is obtained that for the proximity effect with a magnetic monolayer the order parameter is not suppressed, and, on the contrary, is restored to the value in an isolated superconductor at high values of the ferromagnetic exchange field. The physical reasons of this counterintuitive behavior are discussed. The LDOS is also studied as a possible experimental tool for observation of the discussed effects. It is demonstrated that it exhibits clear signatures of the hybridization magnetic proximity effect, what can be observed when a gate voltage is applied to the F layer. The spin splitting of the LDOS cannot be described by an effective Zeeman field  and manifests a multiple peak structure reflecting different effective spin splittings of different superconducting bands. The spin splittings can be tuned by gating. From an applied point of view this opens up the possibility of implementing controlled spin-split superconductivity.

The paper is organized as follows. In Sec.~\ref{sec:model} we formulate a model of the considered heterostructure, and in Sec.~\ref{sec:method} we describe the Green's function approach, which is used further for calculations. In Sec.~\ref{sec:OP} we present results for the dependence of the superconducting order parameter and the LDOS on the number of superconducting layers and discuss the related physics. Sec.~\ref{sec:DOS} studies the experimental manifestations of the hybridization magnetic proximity effect in the LDOS. Sec.~\ref{sec:conclusions} contains the conclusions from our work.

\section{Model}

\label{sec:model}

\begin{figure}[tb]
	\begin{center}
		\includegraphics[width=75mm]{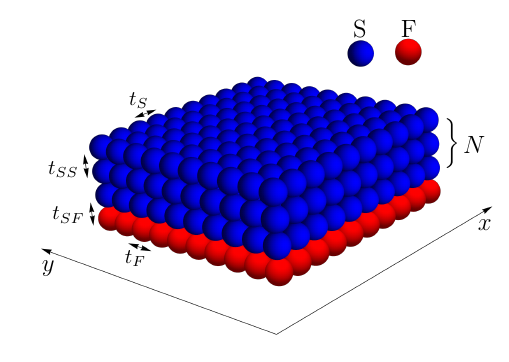}
\caption{{\bf S/F heterostructure under consideration.} It consists of a ferromagnetic monolayer (red) and $N$-layered superconductor (blue). Both materials are modelled by tight-binding Hamiltonians on a square lattice with intralayer hopping elements $t_{S(F)}$ for the S (F) layer, interlayer hopping element $t_{SS}$ between different S layers and hopping element $t_{SF}$ between interface S and F layers.}
 \label{fig:sketch}
	\end{center}
 \end{figure}

The system under consideration is shown in Fig.~\ref{fig:sketch}. It consists of a 2D ferromagnetic layer stacked with N layers of a superconducting material. The system is modelled by the following tight-binding Hamiltonian on a square lattice:
\begin{align}
\hat{H} &= - t_F \sum\limits_{< \bm i \bm j>, \sigma} c^{F+}_{ \bm i \sigma} c^{F}_{\bm j \sigma} - t_S \sum\limits_{n=1}^{N} \sum\limits_{<\bm i \bm j>, \sigma} c^{S(n)+}_{\bm i \sigma} c^{S(n)}_{\bm j \sigma} + \nonumber \\
&\sum\limits_{\bm i}  c^{F+}_{\bm i \alpha} (\bm h \bm \sigma)_{\alpha \beta} c^{F}_{\bm i \beta}- t_{SF} \sum\limits_{\bm i \sigma} (c^{F+}_{\bm i \sigma} c^{S(1)}_{\bm i \sigma} + h.c.) - \nonumber \\
&t_{SS} \sum\limits_{n=1}^{N-1} \sum\limits_{\bm i \sigma} (c^{S(n)+}_{\bm i \sigma} c^{S(n+1)}_{\bm i \sigma} + h.c.) - \nonumber \\
&\mu_{F} \sum\limits_{\bm i \sigma} c^{F+}_{\bm i \sigma} c^{F}_{\bm i \sigma} - \mu_S \sum\limits_{n=1}^{N} \sum\limits_{\bm i \sigma} c^{S(n)+}_{\bm i \sigma} c^{S(n)}_{\bm i \sigma} + \nonumber \\
&\sum\limits_{n=1}^{N} \sum\limits_{\bm i} (\Delta^{(n)}  c^{S(n)+}_{\bm i \uparrow} c^{S(n)+}_{\bm i \downarrow} + h.c.) 
\label{eq:hamiltonian}
\end{align}
Here $c_{\bm i,\sigma}^F$ ($c_{\bm i,\sigma}^{S(n)}$) is an annihilation operator for electrons belonging to the F layer ($n$-th S layer with $n \in [1,N]$) at site $\bm i$ in plane of the layer and for spin $\sigma = \uparrow, \downarrow$. $\mu_{S,F}$ are chemical potentials of the S and F materials respectively, counted from the bottom of the corresponding conduction band. We assume only nearest-neighbor hopping.  $t_{S,F}$ are the nearest-neighbor hopping elements in the planes of the S and F materials. $\langle \bm i \bm j \rangle$ means summation over nearest neighbors. $t_{SF}$ is the hopping element between the S and F layers, and $t_{SS}$ is the hopping element between the different S layers. $\bm h$ is the exchange field of the ferromagnet, which is assumed to be spatially homogeneous in plane of the F layer. $h.c.$ means hermitian conjugate. $\Delta^{(n)}$ is the superconducting order parameter in the superconductor, which is to be calculated self-consistently as $\Delta^{(n)} = \lambda \langle c^{S(n)}_{\bm i \downarrow} c^{S(n)}_{\bm i \uparrow} \rangle$, where $\lambda$ is the pairing constant. Due to the translational invariance of the system along the S/F interface $\Delta^{(n)}$ is also spatially homogeneous along the interface, but it can depend on the number of the superconducting layer, what is described by the superscript $(n)$. Below in the framework of this model we study the superconducting order parameter and the LDOS. 

We take $t_F/t_S = 1.25$, $t_{SS}/t_S = 0.08$ and $t_{SF}/t_S = 0.02 \div 0.04$. This choice of parameters correctly takes into account the main qualitative features of the electronic spectra of vdW materials near the Fermi surface. It follows from results of DFT calculations \cite{Wang2021}, for example, that the interlayer hopping in $\mathrm{NbSe_2}$ is approximately an order of magnitude smaller than the intralayer one, furthermore it is reasonable to assume the hopping between two interface layers of different materials even smaller \cite{Bobkov2024_vdW}, which takes into account the lattice mismatch and possible defects of the interface. The model we study is quite simple and does not contain some essential features of real vdW materials, for example Ising-type spin-orbit coupling and Rashba-type spin-orbit coupling. It can be important for many physical effects, for example, for appearance of finite-momentum pairing (Fulde-Ferrel-Larkin-Ovchnnikov state) \cite{Kaur2005,Akbari2022,Zhang2022,Zhao2023,Wan2023,Ding2023} or dissipationless spin transport \cite{Bobkov2024_spin}. However, when we are interested in magnetic proximity and its influence on the superconducting order parameter and superconducting gap in the LDOS, the considered model captures the main physics determined by the interface hybridization of the electronic spectra of  proximitized materials, as  was demonstrated in the monolayer limit \cite{Bobkov2024_vdW}. In this work we only consider ballistic limit. The influence of impurity scattering on the discussed effects is a prospect for future work.

\section{Green's functions technique for multi-layered S/F vdW heterostructures}

\label{sec:method}

For calculations of the superconducting order parameter and the LDOS we use the Green's functions technique. Since we consider layer-dependent creation and annihilation electron operators and have $N+1$ different layers, the Matsubara Green's function is a $4 (N+1) \times 4 (N+1)$ matrix in the direct product of spin, particle-hole, and layer spaces. Introducing the Nambu spinor \begin{widetext}
$\check \psi_{\bm i} = (c_{{\bm i}\uparrow}^F, c_{\bm i\downarrow}^F, c_{\bm i\uparrow}^{F\dagger}, c_{\bm i\downarrow}^{F\dagger}, c_{{\bm i}\uparrow}^{S(1)}, c_{\bm i\downarrow}^{S(1)}, c_{\bm i\uparrow}^{S(1)\dagger}, c_{\bm i\downarrow}^{S(1)\dagger},..., c_{{\bm i}\uparrow}^{S(N)}, c_{\bm i\downarrow}^{S(N)}, c_{\bm i\uparrow}^{S(N)\dagger}, c_{\bm i\downarrow}^{S(N)\dagger})^T$
\end{widetext}
we define the Green's function as follows: 
\begin{eqnarray}
\check G_{\bm i \bm j}(\tau_1, \tau_2) = - \langle T_\tau \check \psi_{\bm i}(\tau_1) \check \psi_{\bm j}^\dagger(\tau_2) \rangle 
\label{Green_Gorkov}
\end{eqnarray}
where $\langle T_\tau ... \rangle$ means  imaginary time-ordered thermal averaging. The system under consideration is translation invariant along the S/F interface. For this reason one can introduce the Fourier transform of the Green's function:
\begin{eqnarray}
\check G(\bm p, \tau) =  \int d^2 r e^{-i \bm p(\bm i - \bm j)}\check G_{\bm i \bm j},
\label{mixed}
\end{eqnarray}
where the integration is over $\bm i - \bm j$ and $\tau = \tau_1 - \tau_2$.
Introducing Pauli matrices $\sigma_k$ and $\tau_k$ ($k=0,x,y,z$) in spin and particle-hole spaces, respectively, and expanding the Green's function $\check G(\bm p, \tau) $ over fermionic Matsubara frequencies $\omega_m = \pi T(2m+1)$ as $\check G(\bm p, \tau) = T \sum \limits_{\omega_m} e^{-i \omega_m \tau} \check G(\bm p, \omega_m)$, where $T$ is the temperature, 
one can obtain the Gor'kov equation for Green's function. We also define the following  transformed Green's function to simplify further calculations and to present the Gor'kov equation in a more common form:
\begin{eqnarray}
\check {\tilde G}(\bm p,\omega_m) = 
\left(
\begin{array}{cc}
1 & 0 \\
0 & -i\sigma_y
\end{array}
\right)_\tau  \check G(\bm p, \omega_m)  
\left(
\begin{array}{cc}
1 & 0 \\
0 & -i\sigma_y
\end{array}
\right)_\tau ,
\label{unitary}
\end{eqnarray}
where subscript $\tau$ means that the explicit matrix structure corresponds to the particle-hole space. Then we obtain the following Gor'kov equation for $\check {\tilde G}(\bm p,\omega_m)$ (see the Appendix for details of the derivation):
\begin{widetext}
\begin{align}
    \check G^{-1} \check {\tilde G}(\bm p, \omega_m)=1,
\label{Gor'kov_equation}
\end{align}
\begin{equation}
\check{G}^{-1} = \left(\begin{matrix} i \omega_m \tau_z -\zeta_F - \bm h \bm \sigma \tau_z & t_{SF} &0 & 0 & 0 & ...\\ t_{SF} & i \omega_m \tau_z -\zeta_S + i \Delta^{(1)} \tau_y &  t_{SS} & 0 & 0 & ...\\  0&t_{SS}&i \omega_m \tau_z -\zeta_S + i \Delta^{(2)} \tau_y  & t_{SS} & 0 & ... \\  0 & 0 &t_{SS}&i \omega_m \tau_z -\zeta_S + i \Delta^{(3)} \tau_y  & t_{SS} &...\\ ...&...&...&...&...&...\\\end{matrix}\right)
\label{matrix_hamiltonian}
\end{equation}
\end{widetext}
where $\zeta_{S(F)} = -2t_{S(F)}(\cos p_x a+\cos p_y a) - \mu_{S(F)}$ is the normal state electron spectrum of the S(F) layer, and $a$ is the lattice constant. Each element of Eq.~(\ref{matrix_hamiltonian}) is a $4 \times 4$ matrix in the direct product of particle-hole and spin spaces, and the direct structure of this operator in the layer space is written explicitly.

The superconducting order parameter, which can be chosen as a real quantity, is calculated from the self-consistency equation
\begin{eqnarray}
\Delta^{(n)} =  \lambda  T \sum \limits_{\omega_m}\int \frac{d^2 p}{(2\pi)^2} \frac{{\rm Tr}[\check {\tilde G}^{S(n)S(n)}(\bm p, \omega_m)\sigma_0\tau_-]}{4} , ~~
\label{SC}    
\end{eqnarray}
where $\tau_- = \tau_x - i \tau_y$ and $\check {\tilde G}^{S(n)S(n)}$ is a $4 \times 4$ matrix in the direct product of particle-hole and spin spaces, which represents the $(n+1,n+1)$-element of $\check {\tilde G}(\bm p, \omega_m)$ in the layer space, that is, the Green's function of the $n$-th superconducting layer.

The electron spectral function in the $n$-th S-layer for a given spin $\sigma$ is calculated as
\begin{align}
&A^{(n)}_{\sigma} (\varepsilon, \bm p) =-\frac{1}{\pi} \times \nonumber \\
&{\rm Im}[\frac{{\rm Tr}[\check {\tilde G}^{S(n)S(n),R}(\sigma_0+\sigma\sigma_z)(\tau_0+\tau_z)]}{4}],~~~~
\label{DOS}    
\end{align}
$\check {\tilde G}^{S(n)S(n),R}$ can be obtained from $\check {\tilde G}^{S(n)S(n)}$ with the substitution $i\omega_m \to \varepsilon+i\delta$, where $\delta$ is a positive infinitesimal imaginary part. Spin-resolved LDOS in the $n$-th S-layer is calculated as the momentum-integrated spectral function:
\begin{eqnarray}
N^{(n)}_{\sigma}(\varepsilon) = \int \frac{d^2 p}{(2\pi)^2} A^{(n)}_{\sigma}(\varepsilon, \bm p) .
\label{DOS}    
\end{eqnarray}

\section{Dependence of the superconducting order parameter on the number of superconducting layers}

\label{sec:OP}

\begin{figure}[tb]
	\begin{center}
		\includegraphics[width=85mm]{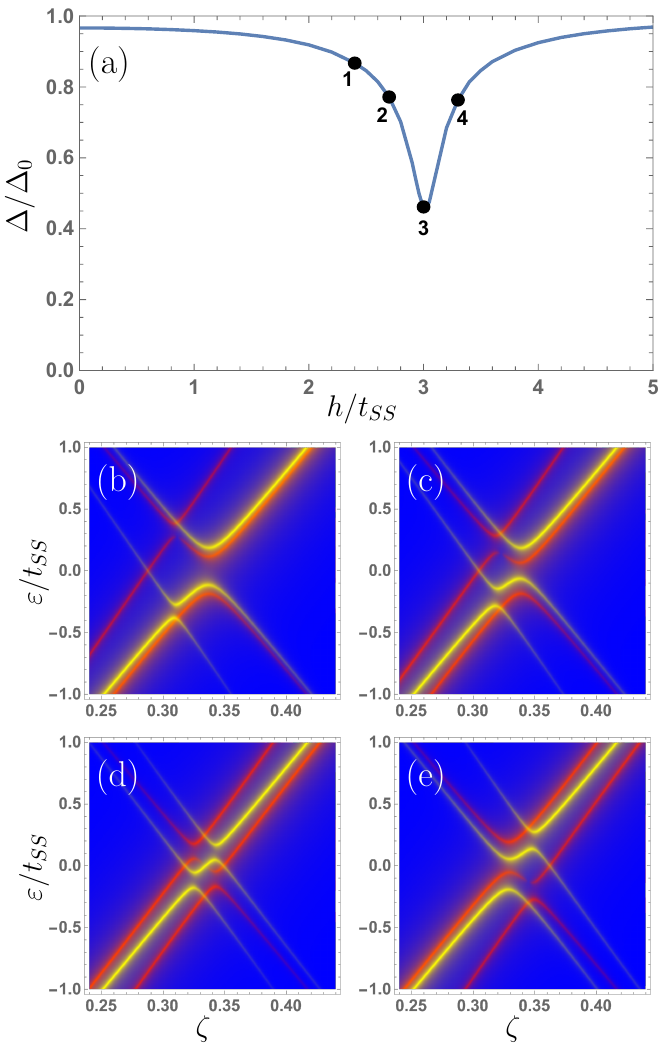}
\caption{{\bf Superconducting order parameter (OP) and electronic spectra for S/F heterostructure with $\bm {N=1}$ S-layer.} (a) Superconducting OP as a function of the exchange field $h$ in the ferromagnet. $\Delta_0$ is the OP of the isolated S-layer. (b)-(e) Electronic spin-up (yellow) and spin-down (red) spectral function as a function of quasiparticles energy $\varepsilon$ and $\zeta = -2(\cos p_x a + \cos p_y a)$ at given values of $h$ denoted by points (b) 1, (c) 2, (d) 3 and (e) 4 in panel (a). $t_S = 12 t_{SS}$, $t_F = 15 t_{SS}$, $t_{SF} = 0.23 t_{SS}$, $\mu_S = 4 t_{SS}$, $\mu_F = 2 t_{SS}$, $T=0.04 t_{SS}$, $\Delta_0 = 0.184 t_{SS}$.}
 \label{fig:vdW_N1}
	\end{center}
 \end{figure}

In this section we discuss how the magnetic proximity evolves with changing in the number of superconducting layers in the system. The hybridization proximity effect for a bilayer S/F heterostructure containing $N=1$ superconducting layer and 1 ferromagnetic layer has already been studied in Ref.~\onlinecite{Bobkov2024_vdW}. Here we describe again the main physics of this proximity effect. In the present paper it serves as a starting point for discussing the dependence of the hybridization proximity effect on the number of layers. We assume that the exchange field of the ferromagnet is directed along the $z$-axis, that is $\bm h = h \hat {\bm z}$. The dependence of the superconducting order parameter (OP) on $h$ is presented in Fig.~\ref{fig:vdW_N1}(a). Only positive values of $h$ are shown, but the plot is symmetric with respect to $h \to -h$. The dependence is nonmonotonic with  only one dip and it is the typical behavior of $\Delta(h)$ for the considered simple tight-binding model on the square lattice. 

The physical mechanism of such a one dip-behavior is illustrated in Figs.~\ref{fig:vdW_N1}(b)-(e), which demonstrate the electronic spectra $\varepsilon(\zeta)$ of the bilayer. More precisely, in Figs.~\ref{fig:vdW_N1}(b)-(e) the electron spectral functions $A_{\uparrow}^{(1)}$ (yellow) and $A_{\downarrow}^{(1)}$ (red) are plotted simultaneously. In the considered model the electronic spectra depend on the momentum only via $\zeta = -2(\cos p_x a + \cos p_y a)$, where $a$ is a lattice constant. For this reason the electronic spectral functions are shown as functions of $\zeta$ instead of more convenient for first-principle calculations representation as a function of the momentum along certain directions in the Brillouin zone (BZ). Figs.~\ref{fig:vdW_N1}(b)-(e) are plotted at the particular values of $h$ corresponding to points 1-4 in Fig.~\ref{fig:vdW_N1}(a), respectively. In Fig.~\ref{fig:vdW_N1}(b) a linear  electronic branch originating from the F layer with a positive slope (red, spin-down) and hyperbolic BCS-like electronic branches coming from the S layer are clearly seen. The second electronic branch of the spin-split spectrum of the F layer with a positive slope (yellow, spin-up) is outside the figure due to the sufficiently large value of the spin splitting $h$. The S-branches are also spin split and this is a signature of their hybridization with the spin-down F-branch \cite{Bobkov2024_vdW} and, consequently, the presence of $h_{eff}$ in the superconductor. The hybridization of the F- and S-branches also manifests itself in the form of anti-crossing at the points of intersection of the branches belonging to the same spin. These anti-crossings result in the finite-energy gaps. In this figure, one can also see weaker linear branches of the spectrum with a negative slope and anti-crossing regions. These are the hole branches of the F layer, which we see in the electron spectral function due to the proximity effect with the superconductor and for this reason the intensity of these lines is small compared to the electron spectral lines due to the smallness of the superconducting OP.

Fig.~\ref{fig:vdW_N1}(c) corresponds to a higher value of $h$. It results in the shift of the spin-down F-branch closer to the S-branches and, consequently, stronger hybridization between the F- and S-branches and higher spin splitting of the S-branches. The OP is more strongly suppressed since the effect of both suppressive factors - $h_{eff}$ induced in the superconductor and the leakage of superconducting correlations into the F layer - increases with increasing degree of hybridization.  

Fig.~\ref{fig:vdW_N1}(d) represents the case of the strongest hybridization between the F- and S-branches, when  it is not even possible to clearly identify F- and S-branches. As a result, the suppression of the OP is maximal. The position of the dip in the dependence $\Delta(h)$ can be found analytically from the condition that, in the absence of superconductivity $\Delta=0$ and interlayer coupling $t_{SF}$ the F- and S-branches intersect at the Fermi level, that is at $\zeta_F+\sigma h = \zeta_S = 0$. It gives the following position of the dip due to the hybridization of the S-branches with the spin-down F-branch ($\sigma = -1$): $h_{dip,\downarrow}^{N=1} = (t_F/t_S)\mu_S - \mu_F$, which is in agreement with the data in Fig.~\ref{fig:vdW_N1}(a).

Upon further increase of $h$ the right red branch starts moving to the right, see Fig.~\ref{fig:vdW_N1}(e). Now it is the spin-down F-branch. The hybridization becomes weaker, the effective exchange field $h_{eff}$ in the superconductor declines in the absolute value and the superconductivity is partially restored. It is interesting that at $h>h_{dip,\downarrow}^{N=1}$ the sign of the spin splitting in the superconductor (that is, the sign of $h_{eff}$) is reversed.

\begin{figure}[tb]
	\begin{center}
		\includegraphics[width=80mm]{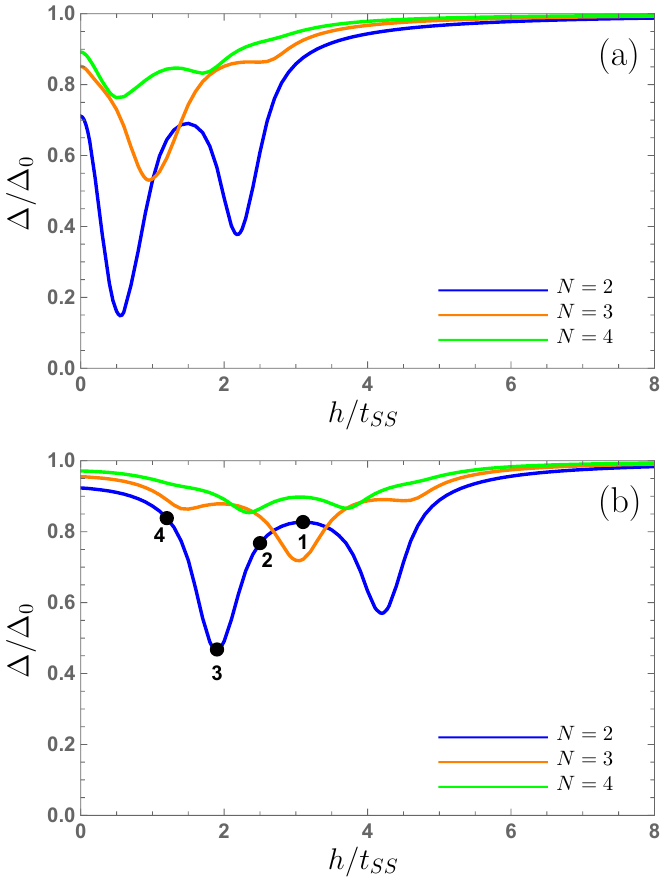}
\caption{{\bf Superconducting order parameter for S/F heterostructure with $\bm{N=2,3,4}$ S-layers.} (a)-(b) Superconducting OP as a function of the exchange field $h$ in the ferromagnet. (a)$\mu_F = 4t_{SS}$, (b)$\mu_F = 2t_{SS}$. $t_{SF} = 0.5 t_{SS}$ for both panels. The other parameters are the same as in Fig.~\ref{fig:vdW_N1}. Points 1-4 in panel (b) correspond to electronic spectra presented in Figs.~\ref{fig:vdW_spectr}(a)-(d), respectively. The OP is calculated self-consistently with the same coupling constant $\lambda$ as for the $N=1$ S/F heterostructure. }
 \label{fig:vdW_N234}
	\end{center}
 \end{figure}

\begin{figure}[tb]
	\begin{center}
		\includegraphics[width=85mm]{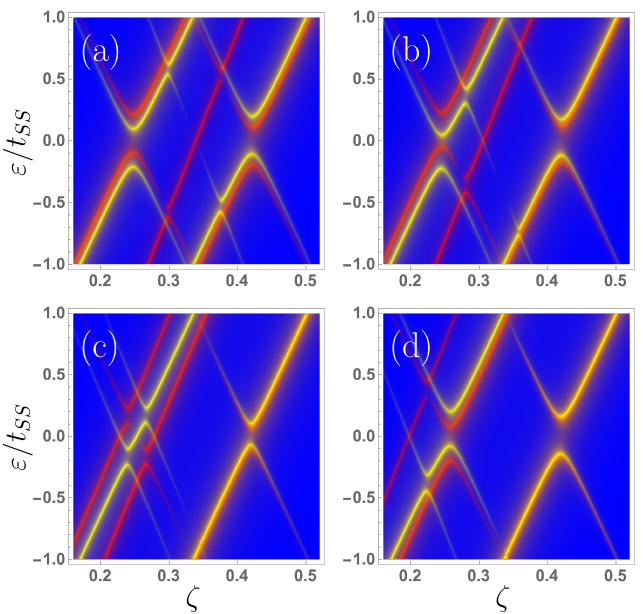}
\caption{{\bf Electronic spectra for S/F heterostructure with $\bm {N=2}$ S-layers.} (a)-(d) Electronic spin-up (yellow) and spin-down (red) spectral function as a function of quasiparticles energy $\varepsilon$ and $\zeta = -2(\cos p_x a + \cos p_y a)$ at given values of $h$ denoted by points (a) 1, (b) 2, (c) 3 and (d) 4 in Fig.~\ref{fig:vdW_N234}. The numerical parameters are the same as in Fig.~\ref{fig:vdW_N234}.}
 \label{fig:vdW_spectr}
	\end{center}
 \end{figure}

For systems with several S-layers self-consistent results for the OP are slightly different for $n=1,2,...,N$ S-layers. The difference is small and the qualitative and quantitative behavior of $\Delta^{(n)}$ is the same for all $n$. For this reason we only present results for the outer S-layer $\Delta^{(n=N)} \equiv \Delta$.

Now let us discuss the S/F bilayer with $N=2$ superconducting layers.  The results for the dependence $\Delta(h)$ are shown by blue curves in Figs.~\ref{fig:vdW_N234}(a) and (b) for two different values of $\mu_F$ corresponding to two physically different situations. In both cases one can see two dips instead of one dip typical for S/F bilayers with $N=1$ S-layer. The reason is that at $N=2$ the electronic spectrum of the S-layer consists of two branches, which are seen in Fig.~\ref{fig:vdW_spectr}. Each of the dips originates from the intersection of one of the F-branches with one of the S-branches giving rise to the highest  possible hybridization degree. Again, approximate positions of the dips can be found analytically. The electronic spectrum of the S-layer in the normal state $\Delta=0$ and in the absence of the coupling to the F-layer $t_{SF}=0$ can be found from the poles of the Green's function determined by Eq.~\ref{matrix_hamiltonian}. In the case $N=2$ it takes the form
\begin{align}
\varepsilon_S^\nu = \zeta_S + \nu t_{SS} ,
\label{eq:spectrum_S2}
\end{align}
where $\nu = \pm 1$. The electronic spectrum of the F-layer takes the form
\begin{align}
\varepsilon_{F,\sigma} = \zeta_F + \sigma h .
\label{eq:spectrum_F2}
\end{align}
From the condition that the intersection of the branches expressed by Eqs.~\ref{eq:spectrum_S2} and \ref{eq:spectrum_F2} occurs at the Fermi surface $\varepsilon_S = \varepsilon_{F,\sigma}=0$ we obtain that the hybridization of one of the S-branches with the F-branch corresponding to spin $\sigma$ takes place at 
\begin{align}
h_{dip,\sigma}^{N=2} = -\sigma [(t_F/t_S)(\mu_S - \nu t_{SS})-\mu_F] .
\label{eq:dips2}
\end{align}
This formula gives the positions of the dips in Figs.~\ref{fig:vdW_N234}(a) and (b) with good accuracy. The small discrepancy between this simple analytical formula and the result of the exact calculation presented in Fig.~\ref{fig:vdW_N234} can be explained by distortion of the spectra due to nonzero superconducting OP and interlayer coupling $t_{SF}$, which were neglected in our analytical consideration. 

Figs.~\ref{fig:vdW_N234}(a) and (b) differ by the mutual arrangement of the F- and S-branches at $h=0$. For the case presented in Fig.~\ref{fig:vdW_N234}(a) the spin-degenerate F-branches are in the middle between two S-branches. In contrast, for the parameters corresponding to Fig.~\ref{fig:vdW_N234}(b) the F-branches are to the left of both S-branches. This leads to weaker hybridization between the F- and the S-branches at $h=0$ and weaker suppression of superconductivity in the latter case, as it is seen from comparison of the blue curves in Figs.~\ref{fig:vdW_N234}(a) and (b). Furthermore, in the case presented in Fig.~\ref{fig:vdW_N234}(a) both S-branches are hybridized with spin-up or spin-down F-branches upon increasing of $h$, while in the case presented in Fig.~\ref{fig:vdW_N234}(b) the spin-up F-branch participates in the hybridization to a smaller extent than the spin-down branch when $h$ grows.  For this reason the OP is suppressed more slowly in the latter case.  

\begin{figure}[tb]
	\begin{center}
		\includegraphics[width=85mm]{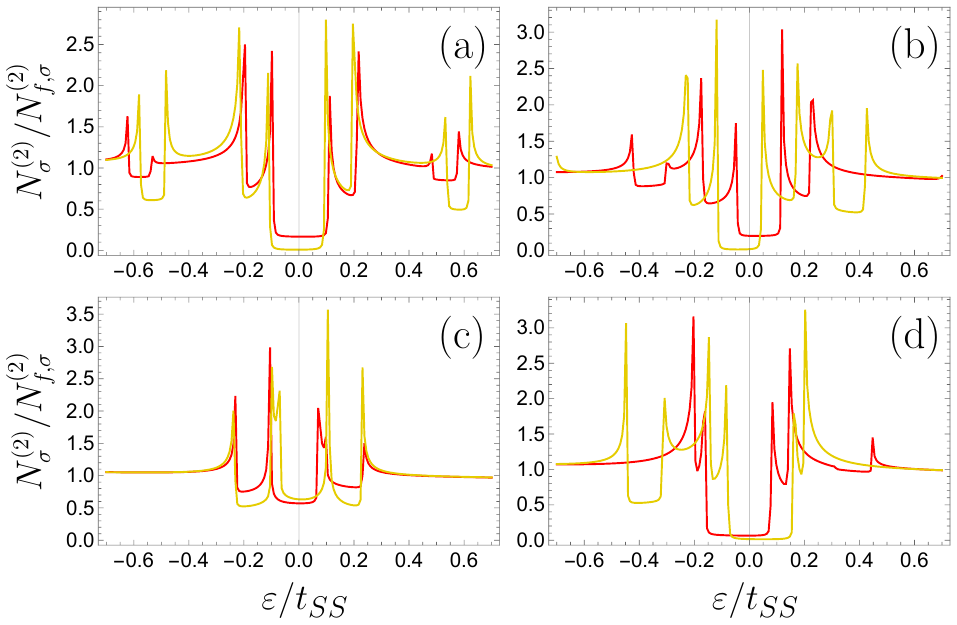}
\caption{{\bf LDOS in the superconductor for the S/F heterostructure with $\bm {N=2}$ S-layers.} Spin-up (yellow) and spin-down (red) LDOS at the outer S-layer $(n=2)$ as a function of the quasiparticle energy. Panels (a)-(d) correspond to different values of $h$ denoted by points (a) 1, (b) 2, (c) 3 and (d) 4 in Fig.~\ref{fig:vdW_N234}. The LDOS is normalized to the normal state LDOS at $T>T_c$ taken at the Fermi level $N_{f,\sigma}^{(2)} = N_\sigma^{(2)}(\varepsilon = \varepsilon_f, T>T_c)$. The numerical parameters are the same as in Fig.~\ref{fig:vdW_N234}.}
 \label{fig:vdW_DOS}
	\end{center}
 \end{figure}

Analogously to the case of the $N=1$ S-layer for $N=2$ system the spin splitting of each of the S-branches changes sign when the exchange field passes through $h_{dip,\sigma}^{N=2}$, which is clearly seen if one compares Figs.~\ref{fig:vdW_spectr}(a)-(b) with Fig.~\ref{fig:vdW_spectr}(d). However, in general the spin splittings of different S-branches are not equal. For this reason the spin splitting of the LDOS, see Eq.~(\ref{DOS}), {\it cannot be described by a single parameter $h_{eff}$ describing the energy shift between the spin-up and spin-down LDOS}. The spin splittings of different S-branches can be approximately found analytically.
Since $\Delta^{(n)}$ are very close at all the S layers, in these calculations for simplicity we put $\Delta^{(n)} = \Delta^{n=N} = \Delta$. The quasiparticle spectra are given by the poles of the Green's function $\check {\tilde G}^R$, that is by zeros of the determinant of $\check G^{-1}$ given by Eq.~(\ref{matrix_hamiltonian}). At $N=1$ the corresponding equation takes the form:
\begin{align}
    \det\left(
\begin{array}{cc}
-\zeta_F-h\sigma \tau_z +\varepsilon\tau_z & t_{SF} \\
t_{SF} & -\zeta_S+\Delta i\tau_y+\varepsilon\tau_z
\end{array}
\right)=0.
\end{align}
Expanding solutions of the above equation up to the leading order in $t_{SF}$ we obtain the following four branches of the electronic spectrum (for a given spin $\sigma$):
\begin{align}
&\varepsilon_S=\pm\sqrt{\zeta_S^2+\Delta^2}+ \nonumber \\
&\frac{(\Delta^2+\zeta_S^2+\zeta_S\zeta_F\mp h\sigma\sqrt{\zeta_S^2+\Delta^2})t_{SF}^2}{\sqrt{\Delta^2+\zeta_S^2}\left((h\sigma\mp \sqrt{\Delta^2+\zeta_S^2})^2-\zeta_F^2 \right)},
\label{spectrum_1_full}
\end{align}
\begin{align}
\varepsilon_F=\pm \zeta_F+h\sigma+t_{SF}^2\frac{\pm h\sigma+\zeta_F+\zeta_S}{\Delta^2+\zeta_S^2-(\pm h\sigma+\zeta_F)^2}.
\end{align}
At $t_{SF} \to 0$ the first two branches represent superconducting branches and the second two branches originate from the ferromagnet. Now let us find the spin splitting of the S branches in the vicinity of the coherence peak, that is  at $\zeta_S=0$. We only write down electronic parts of the S-branches corresponding to the upper sign in Eq.~(\ref{spectrum_1_full}). In this case,
\begin{align}
\varepsilon_S(\zeta_S=0)=\Delta+t_{SF}^2\frac{-h\sigma + \Delta}{(-h\sigma + \Delta)^2-\zeta_F^2}.
\label{spectrum_coherence}
\end{align}
The effective exchange field describing the spin splitting of the S branch can be found as follows:
\begin{align}
    h_{eff}=\frac{\varepsilon_S(\zeta_S=0,\sigma=1)-\varepsilon_S(\zeta_S=0,\sigma=-1)}{2}.
\end{align}
Substituting Eq.~(\ref{spectrum_coherence}) into this definition we obtain $h_{eff}$:
\begin{align}
    h_{eff}=\frac{h(\Delta^2+\zeta_F^2-h^2)t_{SF}^2}{((-h+\Delta)^2-\zeta_F^2)((h+\Delta)^2-\zeta_F^2)}.
\end{align}
Since $\Delta \ll (h,\zeta_F)$, we can simplify the above expression by setting $\Delta=0$, which results in:
\begin{align}
    h_{eff}=\frac{ht_{SF}^2}{\zeta_F^2-h^2}.
    \label{heff_1}
\end{align}
Taking into account that $ \zeta_S=-\mu_S+t_S\zeta = 0$ we obtain that $\zeta_F=-\mu_F+t_F\zeta = -\mu_F + \mu_S (t_F/t_S)$.  Substituting this value of $\zeta_F$ into Eq.~(\ref{heff_1}) we obtain 
\begin{align}
    h_{eff}=\frac{ht_{SF}^2}{h^2-(\mu_S(t_F/t_S)-\mu_F)^2},
    \label{heff_1_final}
\end{align}
which is in a very good agreement with the spin splitting of electronic spectra presented in Figs.~\ref{fig:vdW_N1}(b),(c) and (e). For data shown in  Figs.~\ref{fig:vdW_N1}(d) the hybridization between the F and S branches is maximal, which corresponds to $h = -\sigma(-\mu_F+\frac{t_F}{t_S}\mu_S)$. In this case Eqs.~(\ref{heff_1})-(\ref{heff_1_final}) do not work because the condition $|t_{SF}| \ll \sqrt{h^2 - \zeta_F^2}$ is violated. Under these resonance conditions  $h_{eff}$ reaches its maximal value $h_{eff} \sim t_{SF}$.

At $N=2$ the exact equation for finding all branches of the hybridized spectrum takes the form:
\begin{widetext}
\begin{align}
    \det\left(
\begin{array}{ccc}
-\zeta_F-h\sigma \tau_z +\varepsilon\tau_z & t_{SF} & 0 \\
t_{SF} & -\zeta_S+\Delta i\tau_y+\varepsilon\tau_z & t_{SS} \\
0 & t_{SS} & -\zeta_S+\Delta i\tau_y+\varepsilon\tau_z
\end{array}
\right)=0
\end{align}    

It has 6 solutions, two of which originate from the F layer and other four originate from the S layer. Expanding the solutions originated from the S layer up to second order with respect to $t_{SF}$ we obtain:
\begin{align}
\varepsilon_S^\nu=\pm \sqrt{\Delta^2+(\nu t_{SS}+\zeta_S)^2}+ 
t_{SF}^2\frac{\Delta^2+(\nu t_{SS}+\zeta_S)(\nu t_{SS}+\zeta_F+\zeta_S)\mp h\sigma \sqrt{\Delta^2+(\nu t_{SS}+\zeta_S)^2}}{2\sqrt{\Delta^2+(\nu t_{SS}+\zeta_S)^2}((\pm h \sigma-\sqrt{\Delta^2+(\nu t_{SS}+\zeta_S)^2})^2-\zeta_F^2)}.
\label{spectrum_2_full}
\end{align}
\end{widetext}
In the vicinity of  the coherence peak by setting $\zeta_S + \nu t_{SS}=0$ Eq.~(\ref{spectrum_2_full}) can be simplified as follows (again, only electronic parts of the S-branches are written down):
\begin{align}
\varepsilon_S^\nu=\Delta+\frac{t_{SF}^2}{2}\frac{-h\sigma + \Delta}{(-h\sigma + \Delta)^2-\zeta_F^{\nu 2}},
\label{spectrum_2_coherence}
\end{align}
where
\begin{align}
    \zeta_F^\nu=-\mu_F+\frac{t_F}{t_S}(\nu t_{SS}+\mu_S).
    \label{zeta_F_2}
\end{align}
Analogously to the case $N=1$ we can find the appropriate effective exchange field giving the spin splitting of the spectrum. However, at $N=2$ it is different for different S-branches:
\begin{align}
    h_{eff}^\nu=\frac{\varepsilon_S^\nu(\zeta_S=0,\sigma=1)-\varepsilon_S^\nu(\zeta_S=0,\sigma=-1)}{2}.
    \label{heff_2}
\end{align}
Substituting into Eq.~(\ref{heff_2}) $\varepsilon_S^\nu$ from Eq.~(\ref{spectrum_2_coherence}) we find: 
\begin{align}
    h_{eff}^\nu=\frac{h(\Delta^2+\zeta_F^{\nu 2}-h^2)t_{SF}^2}{2((-h+\Delta)^2-\zeta_F^{\nu 2})((h+\Delta)^2-\zeta_F^{\nu 2})},
\end{align}
where $\zeta_F^\nu$ is expressed by Eq.~(\ref{zeta_F_2}). Further simplifying the above equation by setting $\Delta=0$, we finally obtain:
\begin{align}
    h_{eff}^\nu=\frac{ht_{SF}^2}{2(\zeta_F^{\nu 2}-h^2)}.
    \label{heff_2_final}
\end{align}
As in the case of $N=1$, this expression describes well the spin splitting of the spectra obtained numerically and presented in Figs.~\ref{fig:vdW_spectr}(a),(b) and (d). The data shown in  Fig.~\ref{fig:vdW_spectr}(c) correspond to the resonance condition $h = -\sigma \zeta_F^\nu$ and are not described by Eq.~(\ref{heff_2_final}).

LDOS plotted for the same values of $h$, which correspond to the electronic spectra presented in Figs.~\ref{fig:vdW_spectr}(a)-(d), are shown in Figs.~\ref{fig:vdW_DOS}(a)-(d), respectively. As seen from Fig.~\ref{fig:vdW_spectr}(a), at this value of $h$ the position of the F-branch is nearly symmetric with respect to the S-branches. In this case the spin splittings of the S-branches are nearly the same in  absolute value, but have opposite signs. Both branches equally contribute to the LDOS. This results in the fact that spin-up and spin-down LDOS in Fig.~\ref{fig:vdW_DOS}(a) manifest very close double-split coherence peaks, such that the total LDOS also has the double-split coherence peaks resembling conventional spin splitting of the coherence peaks in Zeeman-split superconductors. However, in the considered case this double splitting is not a result of the opposite energy shifts  of the spin-up and spin-down LDOS due to the Zeeman energy. The spin polarization of the LDOS is small and originates from the F-branch. The presence of the nonzero contribution of the F-branch to the LDOS in the S-layer gives rise to a small non-zero spin-down LDOS inside the gap region.

Moving from Fig.~\ref{fig:vdW_spectr}(a) to Fig.~\ref{fig:vdW_spectr}(b), we see that due to the shift of the F-branch to the left, the antisymmetry of the spin splitting of the two superconducting branches is violated. As a result, in the LDOS shown in Fig.~\ref{fig:vdW_DOS}(b) we see a higher value of spin polarization at all energies, including the regions of coherent peaks. At the same time, as before, both the spin-up and spin-down LDOS contain doubled coherent peaks, which is a consequence of the presence of two S-branches with their own spin splitting. Moving further to Fig.~\ref{fig:vdW_spectr}(c), which represents the case of the strongest hybridization of the F-branch with one of the S-branches we see that the gap is closed for both (spin-up and spin-down) LDOS. At the same time double-split coherence peaks survive. The former feature results from the fact that due to the strong hybridization with the gapless F-branch the left S-branch loses its gap at the Fermi level, and the latter feature originates from the right S-branch, which is distorted by the hybridization only slightly. And, finally, going to Fig.~\ref{fig:vdW_spectr}(d), where the F-branch is located to the left of both S-branches, we see that the Zeeman splittings of both branches have the same sign and, in addition, the right S-branch is weakly affected by hybridization. Consequently, the corresponding LDOS, shown in Fig.~\ref{fig:vdW_DOS}(d), can be approximately described by an effective Zeeman splitting $h_{eff}$, which gives an energy shift of spin-up and spin-down LDOS in opposite directions. At the same time, for the spin-up LDOS, the finite-energy superconducting gap is also clearly visible, in accordance with Fig.~\ref{fig:vdW_spectr}(d).

Now we discuss further evolution of the OP in S/F bilayers with $N>2$. Two typical examples of the behavior $\Delta(h)$ at $N=3$ and $N=4$ are presented in Figs.~\ref{fig:vdW_N234}(a) and (b) by orange and green lines. In case of $N$ S-layers we have $N$ S-branches. Their energies take the form
\begin{align}
\varepsilon_S = \zeta_S + 2 t_{SS} \cos [\frac{\pi m}{N+1}], ~~ m \in [1,...,N] .
\label{eq:spectrum_SN}
\end{align}
In general, for $N$ layers the dependence $\Delta(h)$ manifests $\leq N$ dips. Each of the dips originates from the intersection between the spin-up or spin-down F-branch and one of the S-branches. This situation is shown in Fig.~\ref{fig:vdW_N234}(b) [orange curve]. At the same time, in some special cases when the initial position of the spin-degenerate F-branches at $h=0$ is located symmetrically between two S-branches (this is controlled by the difference between $\mu_F$ and $\mu_S$) upon further increase of $h$ the spin-split F-branches moving in the opposite directions intersect with these S-branches simultaneously. Then two dips resulting from these hybridization events occur at the same $h$ and one observes only one stronger dip at this value of the exchange field. This situation is shown in Fig.~\ref{fig:vdW_N234}(a) (orange curve corresponding to $N=3$). 

Another important thing that we can see in Figs.~\ref{fig:vdW_N234}(a) and (b) is that the depth of the dips becomes smaller with increasing number of the S-layers. It is explained by the fact that we have only one spin-up and only one spin-down F-branches. Each of these branches can be hybridized strongly only with one of S-branches simultaneously. Other S-branches are influenced by the hybridization much weaker. However, all the S-branches contribute to the superconducting OP. Therefore, superconductivity suppression is reduced with increasing  number of S-layers. In the limit of large number of the S-layers the multiple-dip structure is smeared into one wide and not very deep  minimum, the width of which is determined by the total width of the energy band occupied by all S-branches. Although the position of the minimum is determined by the difference $\mu_F-\mu_S$, such a limiting behavior is universal for the considered model with rather simple Fermi-surfaces.  More complicated limiting behavior can occur for superconducting materials with complex Fermi surfaces consisting of different sheets. It is interesting that the obtained limiting behavior $\Delta(h)$ for large number of superconducting layers is not similar to well-known classical picture of monotonic suppression of the superconducting OP by the exchange field of the F-layer in thin-film S/F heterostructures, where both layers contain large number of atomic layers \cite{Sarma1963}. To obtain from our result a further transition to this well-known case, it is necessary to consider a larger number of ferromagnetic layers and assume that the interlayer hopping is
of the same order as the intralayer one, which means that the materials are far from the anisotropic vdW limit. 

To end this section, we would like to note that when taking into account the Ising spin-orbit interaction, which is often present in van der Waals materials with a small number of layers \cite{Xi2016_2,delaBarrera2018,Dvir2018,Sohn2018}, the proximity effects we have considered can become anisotropic.  In particular, the magnitude of the suppression of the superconducting OP and the spin splitting of the LDOS can begin to depend on the direction of the exchange field of the ferromagnet relative to the normal to the S/F interface.

\section{Gate-controlled superconducting gap}

\label{sec:DOS}

\begin{figure}[tb]
	\begin{center}
		\includegraphics[width=85mm]{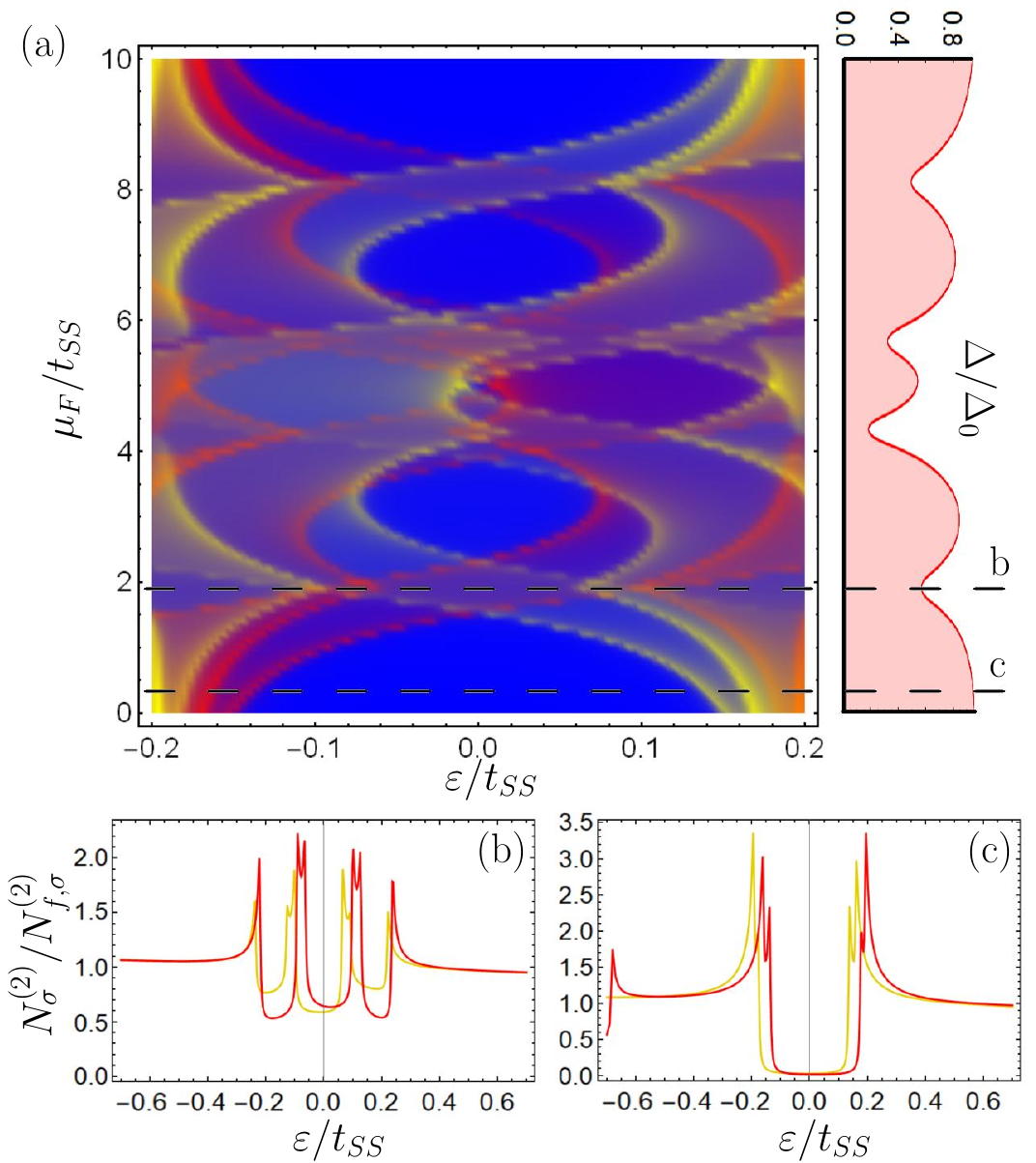}
\caption{{\bf Gate-controlled LDOS and OP in the superconductor for the S/F heterostructure with $\bm {N=2}$ S-layers.} (a) Left panel: spin-up (yellow) and spin-down (red) LDOS at the outer S-layer $(n=2)$ as a function of the quasiparticle energy and chemical potential of the F-layer $\mu_F$. The LDOS is normalized to the normal state LDOS at $T>T_c$ taken at the Fermi level $N_{f,\sigma}^{(2)} = N_\sigma^{(2)}(\varepsilon = \varepsilon_f, T>T_c)$. Right panel: superconducting OP as a function of $\mu_F$. Black dashed lines denoted by letters b and c indicate values of $\mu_F$ chosen for plotting the LDOS in panels (b) and (c), respectively. (b)-(c) Spin-up (yellow) and spin-down (red) LDOS at the outer S-layer $(n=2)$ as a function of the quasiparticle energy. $h=2t_{SS}$. The other numerical parameters are the same as in Fig.~\ref{fig:vdW_N234}.}
 \label{fig:vdW_DOS_gate}
	\end{center}
 \end{figure}

The fundamental physics of the magnetic proximity effect in S/F heterostructures with a vdW magnetic monolayer and a few layer superconductor can be observed in experimental investigations of LDOS via tunneling conductance or by the scanning  tunneling microscopy (STM) technique. In an experiment, it is difficult to vary the exchange field of the F-layer. However, the same physics can be observed when changing the chemical potential of the F-layer using the gate voltage \cite{Xi2016,Deng2018,Matsuoka2023}. Varying the chemical potential also leads to the shift of the F-branches of the spectrum relative to the S-branches. The only difference with the case of varying $h$ is that when the chemical potential is changed, the spin-up and spin-down branches shift in the same direction, and not in the opposite directions. 

Fig.~\ref{fig:vdW_DOS_gate}(a) represents the spin-resolved LDOS for the S/F heterostructure with $N=2$ S-layers, where the spin-up (down) LDOS is shown in yellow (red) color as a function of the quasiparticle energy $\varepsilon$ and the chemical potential of the F-layer $\mu_F$. Blue color corresponds to zero LDOS (gap). The right panel of this figure additionally demonstrates the corresponding superconducting  OP for a given $\mu_F$. By inspecting this dependence one can see that the number of dips is twice  the number of dips in the dependence $\Delta(h)$. It is explained by the fact that upon varying $\mu_F$ spin-up and spin-down F-branches shift in the same direction. Thus, each of the S-branches is hybridized with both F-branches, and not with just one, as is the case when $h$ is varied.

In the dependence of the LDOS on $\mu_F$ presented in Fig.~\ref{fig:vdW_DOS_gate}(a) we can find most of the physical situations discussed in the previous section. Figs.~\ref{fig:vdW_DOS_gate}(b) and (c) demonstrate two examples of the spin-resolved LDOS at given values of $\mu_F$ marked by black dashed lines in Fig.~\ref{fig:vdW_DOS_gate}(a). Results presented in Fig.~\ref{fig:vdW_DOS_gate}(b) correspond to the value of $\mu_F$ providing maximal possible hybridization between F- and S-branches. The plots manifest the same characteristic features as in Fig.~\ref{fig:vdW_DOS}(c): the gap is closed for both (spin-up and spin-down) LDOS, but the double-split coherence peaks survive for both spin subbands. For the case shown in Fig.~\ref{fig:vdW_DOS_gate}(c) the spin-up and spin-down  LDOS are shifted in opposite directions and, consequently, can be approximately described by an effective Zeeman splitting $h_{eff}$. Physically it is close to the situation presented in Fig.~\ref{fig:vdW_DOS}(d), when both S-branches acquire Zeeman splitting of the same sign.

\section{Conclusions}

We have developed a theory of the magnetic proximity effect in vdW S/F heterostructures with different numbers of superconducting layers in proximity with a monolayer ferromagnet. The evolution of the behavior of the superconducting order parameter as a function of the exchange field of the ferromagnet is investigated. It is shown that this dependence has a highly non-monotonic character, which directly reflects the physical mechanism of the proximity effect in such systems: the influence of the superconducting and magnetic orders on each other is determined by the degree of hybridization of the electronic spectra of the materials. The number of minima in the dependence of the order parameter is determined by the number of superconducting layers. The limiting form of the dependence of the order parameter on the exchange field for a large number of superconducting layers is discussed. This dependence also has a non-monotonic character with a smeared minimum. The width of the minimum is determined by the total width of the energy band occupied by all quantized superconducting branches. It is fundamentally different from the well-known monotonic suppression of superconductivity by the exchange field in thin-film heterostructures with a large number of atomic superconducting and ferromagnetic layers. 

The second important manifestation of the magnetic proximity effect - spin splitting of the electronic LDOS - is also investigated. Analogously to the behavior of the superconducting OP this splitting is strongly different from the well-known picture of Zeeman-split superconducting LDOS, which is very similar to the behavior of the bulk superconducting LDOS in the parallel magnetic field and is well-known for thin-film S/F heterostructures with a large number of atomic layers.  In particular, for the case of the considered vdW heterostructure each of the electronic S-branches has its own spin splitting, which can be of opposite sign. It is shown that the discussed nonmonotonic behavior of the superconducting OP and the spin splitting of the LDOS can be observed experimentally when studying the LDOS upon varying the gate potential applied to the F-layer. 

\label{sec:conclusions}

\begin{acknowledgments}
A.S.I., G.A.B. and I.V.B. acknowledge support from the Theoretical Physics and Mathematics Advancement Foundation “BASIS” via the project No. 23-1-1-51-1. The calculations performed for the gate-controlled LDOS were supported by the Russian Science Foundation via the project No. 24-12-00152. The self-consistent calculations of the superconducting order parameter were supported by a grant from the Ministry of Science and Higher Education of the Russian Federation No. 075-15-2024-632.
\end{acknowledgments}

\appendix*

\section{Derivation of the Gor'kov equation for a layered vdW heterostructure}

In this section we present key steps of the derivation of the Gor'kov equation for our system. The Green’s function Eq.~(\ref{Green_Gorkov}) obeys the following equation:

\begin{eqnarray}
\frac{d \check{G}_{\bm i \bm j}}{d \tau_1} = - \delta(\tau_1 -\tau_2) \delta_{\bm i \bm j} - \langle T_{\tau } \frac{d \check{\psi_{\bm i}} (\tau_1)}{d \tau_1} \check{\psi^{\dag}_{\bm j}} (\tau_2) \rangle
\label{Green_derivative}
\end{eqnarray}

For the system described by Hamiltonian Eq.~(\ref{eq:hamiltonian}) the Heisenberg equation of motion for spinor $\check \psi_{\bm i}$ takes the form:

\begin{widetext}
\begin{align}
\frac{d \check{\psi_{\bm i}}}{d \tau} = [\hat{H}, \check{\psi_{\bm i}}]  = \tau_{z} & \hat{M}_{\bm i \bm j}  \check{\psi_{\bm i}}  =
\tau_{z} \left(\begin{matrix} M^{F}_{\bm i \bm j} & t_{SF} &0 & 0 & 0 & ...  \\ t_{SF} &M^{S(1)}_{\bm i \bm j} &  t_{SS} & 0 & 0 & ...\\  0&t_{SS}&M^{S(2)}_{\bm i \bm j}  & t_{SS} & 0 & ... \\  0 & 0 &t_{SS}& M^{S(3)}_{\bm i \bm j} & t_{SS} &...\\ ...&...&...&...&...&...\\\end{matrix}\right) \check{\psi_{\bm i}} \label{Heisenberg_equation} \\
& M^{F}_{\bm i \bm j} = t_F \hat{j} + \mu_F - \bm h \check {\bm \sigma} \nonumber \\
& M^{S(n)}_{\bm i \bm j} = t_S \hat{j} + \mu_S -  \check{\Delta}^{(n)} i\sigma_y , \nonumber
\end{align}
\end{widetext}
where $\check {\bm \sigma} = \bm \sigma (1+\tau_z)/2 + \bm \sigma^* (1-\tau_z)/2$ is the quasiparticle spin operator, $\check{\Delta}^{(n)}=\Delta^{(n)}\tau_{+} + \Delta^{(n)*} \tau_{-}$ with $\tau_{\pm} = (\tau_x\pm i \tau_y)/2$. The operator $\hat{j}$ acts on the spinor $\check{\psi_{\bm i}} $ in the following way:
\begin{equation}
\hat{j} \check{\psi_{\bm i}} = \sum\limits_{\langle ij \rangle} \check{\psi_{\bm i}} = \sum\limits_{\langle \bm a \rangle} \check{\psi}_{\bm i + \bm a}.
\end{equation}
Here $\bm a \in \{\pm \bm a_x, \pm \bm a_y, 0\}$ are the basis vectors in the plane of the layers. Substituting Eq.~ (\ref{Heisenberg_equation}) into Eq.~(\ref{Green_derivative})  we obtain :
\begin{align}
G_{\bm i}^{-1} \check{G}_{\bm i \bm j} & (\tau_1-\tau_2) = \delta_{\bm i \bm j} \delta(\tau_1-\tau_2), \\
& G_{\bm i}^{-1} = \tau_z \hat{M}_{\bm i \bm j} - \frac{d }{d \tau_1}
\end{align}
The operator $\hat{j}$ acts on the Green's function $\check{G}_{\bm i \bm j}$ in the following way:

\begin{equation}
\hat{j} \check{G}_{\bm i \bm j} = \sum\limits_{\bm a} \check{G}_{\bm i + \bm a, \bm j}
\end{equation}
Further we introduce the Fourier transform of the Green’s function as defined by Eq.~(\ref{mixed}). The term $\hat{j} \check{G}_{\bm i \bm j}$ in this mixed representation takes the form:
\begin{eqnarray}
\sum\limits_{\bm a} \int d^2 \bm r e^{-i \bm p (\bm i - \bm j)} \check{G}_{\bm i + \bm a, \bm j} = \nonumber \\
2 \check{G}(\bm p, \tau) [\cos(a_x p_x) + \cos(a_y p_y)],
\end{eqnarray}
where $\tau = \tau_1-\tau_2$. Then, expanding the Green’s function $\check{G}(\bm p, \tau)$ over fermionic Matsubara frequencies, one can obtain the Gor’kov equation for the Green’s function:

\begin{align}
& G_{\bm p}^{-1} (\omega_m) \check{G} (\bm p, \omega_m) = 1, \\
& G_{\bm p}^{-1} (\omega_m) = \tau_z \hat{M}_{\bm p} + i \omega_m,
\end{align}

\begin{small}
\begin{eqnarray}
\hat{M}_{\bm p} =  \left(\begin{matrix} M^{F}_{\bm p} & t_{SF} &0 & 0 & 0 & ...\\ t_{SF} &M^{S(1)}_{\bm p} &  t_{SS} & 0 & 0 & ...\\  0&t_{SS}&M^{S(2)}_{\bm p}  & t_{SS} & 0 & ... \\  0 & 0 &t_{SS}& M^{S(3)}_{\bm p} & t_{SS} &...\\ ...&...&...&...&...&...\\\end{matrix}\right)
\end{eqnarray}
\end{small}
\vspace{-0.4cm}
\begin{eqnarray}
    \begin{aligned}
M^{F}_{\bm p} = 2 t_F [\cos(a_x p_x) + \cos(a_y p_y)] + \\
+ \mu_F 
 - \bm h \check {\bm \sigma} 
     \end{aligned}
\end{eqnarray}
\vspace{-0.4cm}
\begin{eqnarray}
    \begin{aligned}
 M^{S(n)}_{\bm p} = 2 t_S [\cos(a_x p_x) + \cos(a_y p_y)] +\\
 + \mu_S - \check{\Delta}^{(n)} i\sigma_y .
    \end{aligned}
\end{eqnarray}
To simplify further calculations and to present the Gor’kov equation in a more common form, we also define the transformed Green’s function Eq.~(\ref{unitary}). As a result, we obtain the Gor’kov equation Eq.~(\ref{Gor'kov_equation})-(\ref{matrix_hamiltonian}).

\bibliography{vdW_many}

\begin{thebibliography}{59}%
\makeatletter
\providecommand \@ifxundefined [1]{%
 \@ifx{#1\undefined}
}%
\providecommand \@ifnum [1]{%
 \ifnum #1\expandafter \@firstoftwo
 \else \expandafter \@secondoftwo
 \fi
}%
\providecommand \@ifx [1]{%
 \ifx #1\expandafter \@firstoftwo
 \else \expandafter \@secondoftwo
 \fi
}%
\providecommand \natexlab [1]{#1}%
\providecommand \enquote  [1]{``#1''}%
\providecommand \bibnamefont  [1]{#1}%
\providecommand \bibfnamefont [1]{#1}%
\providecommand \citenamefont [1]{#1}%
\providecommand \href@noop [0]{\@secondoftwo}%
\providecommand \href [0]{\begingroup \@sanitize@url \@href}%
\providecommand \@href[1]{\@@startlink{#1}\@@href}%
\providecommand \@@href[1]{\endgroup#1\@@endlink}%
\providecommand \@sanitize@url [0]{\catcode `\\12\catcode `\$12\catcode `\&12\catcode `\#12\catcode `\^12\catcode `\_12\catcode `\%12\relax}%
\providecommand \@@startlink[1]{}%
\providecommand \@@endlink[0]{}%
\providecommand \url  [0]{\begingroup\@sanitize@url \@url }%
\providecommand \@url [1]{\endgroup\@href {#1}{\urlprefix }}%
\providecommand \urlprefix  [0]{URL }%
\providecommand \Eprint [0]{\href }%
\providecommand \doibase [0]{https://doi.org/}%
\providecommand \selectlanguage [0]{\@gobble}%
\providecommand \bibinfo  [0]{\@secondoftwo}%
\providecommand \bibfield  [0]{\@secondoftwo}%
\providecommand \translation [1]{[#1]}%
\providecommand \BibitemOpen [0]{}%
\providecommand \bibitemStop [0]{}%
\providecommand \bibitemNoStop [0]{.\EOS\space}%
\providecommand \EOS [0]{\spacefactor3000\relax}%
\providecommand \BibitemShut  [1]{\csname bibitem#1\endcsname}%
\let\auto@bib@innerbib\@empty
\bibitem [{\citenamefont {Buzdin}(2005)}]{Buzdin2005}%
  \BibitemOpen
  \bibfield  {author} {\bibinfo {author} {\bibfnamefont {A.~I.}\ \bibnamefont {Buzdin}},\ }\bibfield  {title} {\bibinfo {title} {Proximity effects in superconductor-ferromagnet heterostructures},\ }\href {https://doi.org/10.1103/RevModPhys.77.935} {\bibfield  {journal} {\bibinfo  {journal} {Rev. Mod. Phys.}\ }\textbf {\bibinfo {volume} {77}},\ \bibinfo {pages} {935} (\bibinfo {year} {2005})}\BibitemShut {NoStop}%
\bibitem [{\citenamefont {Bergeret}\ \emph {et~al.}(2005)\citenamefont {Bergeret}, \citenamefont {Volkov},\ and\ \citenamefont {Efetov}}]{Bergeret2005}%
  \BibitemOpen
  \bibfield  {author} {\bibinfo {author} {\bibfnamefont {F.~S.}\ \bibnamefont {Bergeret}}, \bibinfo {author} {\bibfnamefont {A.~F.}\ \bibnamefont {Volkov}},\ and\ \bibinfo {author} {\bibfnamefont {K.~B.}\ \bibnamefont {Efetov}},\ }\bibfield  {title} {\bibinfo {title} {Odd triplet superconductivity and related phenomena in superconductor-ferromagnet structures},\ }\href {https://doi.org/10.1103/RevModPhys.77.1321} {\bibfield  {journal} {\bibinfo  {journal} {Rev. Mod. Phys.}\ }\textbf {\bibinfo {volume} {77}},\ \bibinfo {pages} {1321} (\bibinfo {year} {2005})}\BibitemShut {NoStop}%
\bibitem [{\citenamefont {Bergeret}\ \emph {et~al.}(2018)\citenamefont {Bergeret}, \citenamefont {Silaev}, \citenamefont {Virtanen},\ and\ \citenamefont {Heikkil\"a}}]{Bergeret2018review}%
  \BibitemOpen
  \bibfield  {author} {\bibinfo {author} {\bibfnamefont {F.~S.}\ \bibnamefont {Bergeret}}, \bibinfo {author} {\bibfnamefont {M.}~\bibnamefont {Silaev}}, \bibinfo {author} {\bibfnamefont {P.}~\bibnamefont {Virtanen}},\ and\ \bibinfo {author} {\bibfnamefont {T.~T.}\ \bibnamefont {Heikkil\"a}},\ }\bibfield  {title} {\bibinfo {title} {Colloquium: Nonequilibrium effects in superconductors with a spin-splitting field},\ }\href {https://doi.org/10.1103/RevModPhys.90.041001} {\bibfield  {journal} {\bibinfo  {journal} {Rev. Mod. Phys.}\ }\textbf {\bibinfo {volume} {90}},\ \bibinfo {pages} {041001} (\bibinfo {year} {2018})}\BibitemShut {NoStop}%
\bibitem [{\citenamefont {Heikkil{\"a}}\ \emph {et~al.}(2019)\citenamefont {Heikkil{\"a}}, \citenamefont {Silaev}, \citenamefont {Virtanen},\ and\ \citenamefont {Bergeret}}]{Heikkila2019review}%
  \BibitemOpen
  \bibfield  {author} {\bibinfo {author} {\bibfnamefont {T.~T.}\ \bibnamefont {Heikkil{\"a}}}, \bibinfo {author} {\bibfnamefont {M.}~\bibnamefont {Silaev}}, \bibinfo {author} {\bibfnamefont {P.}~\bibnamefont {Virtanen}},\ and\ \bibinfo {author} {\bibfnamefont {F.~S.}\ \bibnamefont {Bergeret}},\ }\bibfield  {title} {\bibinfo {title} {Thermal, electric and spin transport in superconductor/ferromagnetic-insulator structures},\ }\href {https://www.sciencedirect.com/science/article/pii/S0079681619300115} {\bibfield  {journal} {\bibinfo  {journal} {Progress in Surface Science}\ }\textbf {\bibinfo {volume} {94}},\ \bibinfo {pages} {100540} (\bibinfo {year} {2019})}\BibitemShut {NoStop}%
\bibitem [{\citenamefont {Linder}\ and\ \citenamefont {Robinson}(2015)}]{Linder2015}%
  \BibitemOpen
  \bibfield  {author} {\bibinfo {author} {\bibfnamefont {J.}~\bibnamefont {Linder}}\ and\ \bibinfo {author} {\bibfnamefont {J.~W.~A.}\ \bibnamefont {Robinson}},\ }\bibfield  {title} {\bibinfo {title} {Superconducting spintronics},\ }\href {https://doi.org/10.1038/nphys3242} {\bibfield  {journal} {\bibinfo  {journal} {Nature Physics}\ }\textbf {\bibinfo {volume} {11}},\ \bibinfo {pages} {307} (\bibinfo {year} {2015})}\BibitemShut {NoStop}%
\bibitem [{\citenamefont {Eschrig}(2015)}]{Eschrig2015}%
  \BibitemOpen
  \bibfield  {author} {\bibinfo {author} {\bibfnamefont {M.}~\bibnamefont {Eschrig}},\ }\bibfield  {title} {\bibinfo {title} {Spin-polarized supercurrents for spintronics: a review of current progress},\ }\href {https://doi.org/10.1088/0034-4885/78/10/104501} {\bibfield  {journal} {\bibinfo  {journal} {Reports on Progress in Physics}\ }\textbf {\bibinfo {volume} {78}},\ \bibinfo {pages} {104501} (\bibinfo {year} {2015})}\BibitemShut {NoStop}%
\bibitem [{\citenamefont {Machon}\ \emph {et~al.}(2013)\citenamefont {Machon}, \citenamefont {Eschrig},\ and\ \citenamefont {Belzig}}]{Machon2013}%
  \BibitemOpen
  \bibfield  {author} {\bibinfo {author} {\bibfnamefont {P.}~\bibnamefont {Machon}}, \bibinfo {author} {\bibfnamefont {M.}~\bibnamefont {Eschrig}},\ and\ \bibinfo {author} {\bibfnamefont {W.}~\bibnamefont {Belzig}},\ }\bibfield  {title} {\bibinfo {title} {Nonlocal thermoelectric effects and nonlocal onsager relations in a three-terminal proximity-coupled superconductor-ferromagnet device},\ }\href {https://doi.org/10.1103/PhysRevLett.110.047002} {\bibfield  {journal} {\bibinfo  {journal} {Phys. Rev. Lett.}\ }\textbf {\bibinfo {volume} {110}},\ \bibinfo {pages} {047002} (\bibinfo {year} {2013})}\BibitemShut {NoStop}%
\bibitem [{\citenamefont {Ozaeta}\ \emph {et~al.}(2014)\citenamefont {Ozaeta}, \citenamefont {Virtanen}, \citenamefont {Bergeret},\ and\ \citenamefont {Heikkil\"a}}]{Ozaeta2014}%
  \BibitemOpen
  \bibfield  {author} {\bibinfo {author} {\bibfnamefont {A.}~\bibnamefont {Ozaeta}}, \bibinfo {author} {\bibfnamefont {P.}~\bibnamefont {Virtanen}}, \bibinfo {author} {\bibfnamefont {F.~S.}\ \bibnamefont {Bergeret}},\ and\ \bibinfo {author} {\bibfnamefont {T.~T.}\ \bibnamefont {Heikkil\"a}},\ }\bibfield  {title} {\bibinfo {title} {Predicted very large thermoelectric effect in ferromagnet-superconductor junctions in the presence of a spin-splitting magnetic field},\ }\href {https://doi.org/10.1103/PhysRevLett.112.057001} {\bibfield  {journal} {\bibinfo  {journal} {Phys. Rev. Lett.}\ }\textbf {\bibinfo {volume} {112}},\ \bibinfo {pages} {057001} (\bibinfo {year} {2014})}\BibitemShut {NoStop}%
\bibitem [{\citenamefont {Kolenda}\ \emph {et~al.}(2016{\natexlab{a}})\citenamefont {Kolenda}, \citenamefont {Wolf},\ and\ \citenamefont {Beckmann}}]{Kolenda2016}%
  \BibitemOpen
  \bibfield  {author} {\bibinfo {author} {\bibfnamefont {S.}~\bibnamefont {Kolenda}}, \bibinfo {author} {\bibfnamefont {M.~J.}\ \bibnamefont {Wolf}},\ and\ \bibinfo {author} {\bibfnamefont {D.}~\bibnamefont {Beckmann}},\ }\bibfield  {title} {\bibinfo {title} {Observation of thermoelectric currents in high-field superconductor-ferromagnet tunnel junctions},\ }\href {https://doi.org/10.1103/PhysRevLett.116.097001} {\bibfield  {journal} {\bibinfo  {journal} {Phys. Rev. Lett.}\ }\textbf {\bibinfo {volume} {116}},\ \bibinfo {pages} {097001} (\bibinfo {year} {2016}{\natexlab{a}})}\BibitemShut {NoStop}%
\bibitem [{\citenamefont {Kolenda}\ \emph {et~al.}(2016{\natexlab{b}})\citenamefont {Kolenda}, \citenamefont {Machon}, \citenamefont {Beckmann},\ and\ \citenamefont {Belzig}}]{Kolenda2016_2}%
  \BibitemOpen
  \bibfield  {author} {\bibinfo {author} {\bibfnamefont {S.}~\bibnamefont {Kolenda}}, \bibinfo {author} {\bibfnamefont {P.}~\bibnamefont {Machon}}, \bibinfo {author} {\bibfnamefont {D.}~\bibnamefont {Beckmann}},\ and\ \bibinfo {author} {\bibfnamefont {W.}~\bibnamefont {Belzig}},\ }\bibfield  {title} {\bibinfo {title} {Nonlinear thermoelectric effects in high-field superconductor-ferromagnet tunnel junctions},\ }\href {https://doi.org/10.3762/bjnano.7.152} {\bibfield  {journal} {\bibinfo  {journal} {Beilstein J. Nanotechnol.}\ }\textbf {\bibinfo {volume} {7}},\ \bibinfo {pages} {1579} (\bibinfo {year} {2016}{\natexlab{b}})}\BibitemShut {NoStop}%
\bibitem [{\citenamefont {Giazotto}\ \emph {et~al.}(2014)\citenamefont {Giazotto}, \citenamefont {Robinson}, \citenamefont {Moodera},\ and\ \citenamefont {Bergeret}}]{Giazotto2014}%
  \BibitemOpen
  \bibfield  {author} {\bibinfo {author} {\bibfnamefont {F.}~\bibnamefont {Giazotto}}, \bibinfo {author} {\bibfnamefont {J.~W.~A.}\ \bibnamefont {Robinson}}, \bibinfo {author} {\bibfnamefont {J.~S.}\ \bibnamefont {Moodera}},\ and\ \bibinfo {author} {\bibfnamefont {F.~S.}\ \bibnamefont {Bergeret}},\ }\bibfield  {title} {\bibinfo {title} {Proposal for a phase-coherent thermoelectric transistor},\ }\href {https://doi.org/10.1063/1.4893443} {\bibfield  {journal} {\bibinfo  {journal} {Applied Physics Letters}\ }\textbf {\bibinfo {volume} {105}},\ \bibinfo {pages} {062602} (\bibinfo {year} {2014})}\BibitemShut {NoStop}%
\bibitem [{\citenamefont {Giazotto}\ \emph {et~al.}(2015{\natexlab{a}})\citenamefont {Giazotto}, \citenamefont {Heikkil\"a},\ and\ \citenamefont {Bergeret}}]{Giazotto2015}%
  \BibitemOpen
  \bibfield  {author} {\bibinfo {author} {\bibfnamefont {F.}~\bibnamefont {Giazotto}}, \bibinfo {author} {\bibfnamefont {T.~T.}\ \bibnamefont {Heikkil\"a}},\ and\ \bibinfo {author} {\bibfnamefont {F.~S.}\ \bibnamefont {Bergeret}},\ }\bibfield  {title} {\bibinfo {title} {Very large thermophase in ferromagnetic josephson junctions},\ }\href {https://doi.org/10.1103/PhysRevLett.114.067001} {\bibfield  {journal} {\bibinfo  {journal} {Phys. Rev. Lett.}\ }\textbf {\bibinfo {volume} {114}},\ \bibinfo {pages} {067001} (\bibinfo {year} {2015}{\natexlab{a}})}\BibitemShut {NoStop}%
\bibitem [{\citenamefont {Machon}\ \emph {et~al.}(2014)\citenamefont {Machon}, \citenamefont {Eschrig},\ and\ \citenamefont {Belzig}}]{Machon2014}%
  \BibitemOpen
  \bibfield  {author} {\bibinfo {author} {\bibfnamefont {P.}~\bibnamefont {Machon}}, \bibinfo {author} {\bibfnamefont {M.}~\bibnamefont {Eschrig}},\ and\ \bibinfo {author} {\bibfnamefont {W.}~\bibnamefont {Belzig}},\ }\bibfield  {title} {\bibinfo {title} {Giant thermoelectric effects in a proximity-coupled superconductor--ferromagnet device},\ }\href {https://doi.org/10.1088/1367-2630/16/7/073002} {\bibfield  {journal} {\bibinfo  {journal} {New Journal of Physics}\ }\textbf {\bibinfo {volume} {16}},\ \bibinfo {pages} {073002} (\bibinfo {year} {2014})}\BibitemShut {NoStop}%
\bibitem [{\citenamefont {Kolenda}\ \emph {et~al.}(2017)\citenamefont {Kolenda}, \citenamefont {S\"urgers}, \citenamefont {Fischer},\ and\ \citenamefont {Beckmann}}]{Kolenda2017}%
  \BibitemOpen
  \bibfield  {author} {\bibinfo {author} {\bibfnamefont {S.}~\bibnamefont {Kolenda}}, \bibinfo {author} {\bibfnamefont {C.}~\bibnamefont {S\"urgers}}, \bibinfo {author} {\bibfnamefont {G.}~\bibnamefont {Fischer}},\ and\ \bibinfo {author} {\bibfnamefont {D.}~\bibnamefont {Beckmann}},\ }\bibfield  {title} {\bibinfo {title} {Thermoelectric effects in superconductor-ferromagnet tunnel junctions on europium sulfide},\ }\href {https://doi.org/10.1103/PhysRevB.95.224505} {\bibfield  {journal} {\bibinfo  {journal} {Phys. Rev. B}\ }\textbf {\bibinfo {volume} {95}},\ \bibinfo {pages} {224505} (\bibinfo {year} {2017})}\BibitemShut {NoStop}%
\bibitem [{\citenamefont {Rezaei}\ \emph {et~al.}(2018)\citenamefont {Rezaei}, \citenamefont {Kamra}, \citenamefont {Machon},\ and\ \citenamefont {Belzig}}]{Rezaei2018}%
  \BibitemOpen
  \bibfield  {author} {\bibinfo {author} {\bibfnamefont {A.}~\bibnamefont {Rezaei}}, \bibinfo {author} {\bibfnamefont {A.}~\bibnamefont {Kamra}}, \bibinfo {author} {\bibfnamefont {P.}~\bibnamefont {Machon}},\ and\ \bibinfo {author} {\bibfnamefont {W.}~\bibnamefont {Belzig}},\ }\bibfield  {title} {\bibinfo {title} {Spin-flip enhanced thermoelectricity in superconductor-ferromagnet bilayers},\ }\href {https://doi.org/10.1088/1367-2630/aad2a3} {\bibfield  {journal} {\bibinfo  {journal} {New Journal of Physics}\ }\textbf {\bibinfo {volume} {20}},\ \bibinfo {pages} {073034} (\bibinfo {year} {2018})}\BibitemShut {NoStop}%
\bibitem [{\citenamefont {Linder}\ and\ \citenamefont {Bathen}(2016)}]{Linder2016}%
  \BibitemOpen
  \bibfield  {author} {\bibinfo {author} {\bibfnamefont {J.}~\bibnamefont {Linder}}\ and\ \bibinfo {author} {\bibfnamefont {M.~E.}\ \bibnamefont {Bathen}},\ }\bibfield  {title} {\bibinfo {title} {Spin caloritronics with superconductors: Enhanced thermoelectric effects, generalized onsager response-matrix, and thermal spin currents},\ }\href {https://doi.org/10.1103/PhysRevB.93.224509} {\bibfield  {journal} {\bibinfo  {journal} {Phys. Rev. B}\ }\textbf {\bibinfo {volume} {93}},\ \bibinfo {pages} {224509} (\bibinfo {year} {2016})}\BibitemShut {NoStop}%
\bibitem [{\citenamefont {Bobkova}\ and\ \citenamefont {Bobkov}(2017)}]{Bobkova2017}%
  \BibitemOpen
  \bibfield  {author} {\bibinfo {author} {\bibfnamefont {I.~V.}\ \bibnamefont {Bobkova}}\ and\ \bibinfo {author} {\bibfnamefont {A.~M.}\ \bibnamefont {Bobkov}},\ }\bibfield  {title} {\bibinfo {title} {Thermospin effects in superconducting heterostructures},\ }\href {https://doi.org/10.1103/PhysRevB.96.104515} {\bibfield  {journal} {\bibinfo  {journal} {Phys. Rev. B}\ }\textbf {\bibinfo {volume} {96}},\ \bibinfo {pages} {104515} (\bibinfo {year} {2017})}\BibitemShut {NoStop}%
\bibitem [{\citenamefont {Bobkova}\ \emph {et~al.}(2021)\citenamefont {Bobkova}, \citenamefont {Bobkov},\ and\ \citenamefont {Belzig}}]{Bobkova2021}%
  \BibitemOpen
  \bibfield  {author} {\bibinfo {author} {\bibfnamefont {I.~V.}\ \bibnamefont {Bobkova}}, \bibinfo {author} {\bibfnamefont {A.~M.}\ \bibnamefont {Bobkov}},\ and\ \bibinfo {author} {\bibfnamefont {W.}~\bibnamefont {Belzig}},\ }\bibfield  {title} {\bibinfo {title} {Thermally induced spin-transfer torques in superconductor/ferromagnet bilayers},\ }\href {https://doi.org/10.1103/PhysRevB.103.L020503} {\bibfield  {journal} {\bibinfo  {journal} {Phys. Rev. B}\ }\textbf {\bibinfo {volume} {103}},\ \bibinfo {pages} {L020503} (\bibinfo {year} {2021})}\BibitemShut {NoStop}%
\bibitem [{\citenamefont {Bobkov}\ \emph {et~al.}(2021)\citenamefont {Bobkov}, \citenamefont {Bobkova}, \citenamefont {Bobkov},\ and\ \citenamefont {Kamra}}]{Bobkov2021}%
  \BibitemOpen
  \bibfield  {author} {\bibinfo {author} {\bibfnamefont {G.~A.}\ \bibnamefont {Bobkov}}, \bibinfo {author} {\bibfnamefont {I.~V.}\ \bibnamefont {Bobkova}}, \bibinfo {author} {\bibfnamefont {A.~M.}\ \bibnamefont {Bobkov}},\ and\ \bibinfo {author} {\bibfnamefont {A.}~\bibnamefont {Kamra}},\ }\bibfield  {title} {\bibinfo {title} {Thermally induced spin torque and domain-wall motion in superconductor/antiferromagnetic-insulator bilayers},\ }\href {https://doi.org/10.1103/PhysRevB.103.094506} {\bibfield  {journal} {\bibinfo  {journal} {Phys. Rev. B}\ }\textbf {\bibinfo {volume} {103}},\ \bibinfo {pages} {094506} (\bibinfo {year} {2021})}\BibitemShut {NoStop}%
\bibitem [{\citenamefont {Giazotto}\ \emph {et~al.}(2006{\natexlab{a}})\citenamefont {Giazotto}, \citenamefont {Heikkil\"a}, \citenamefont {Luukanen}, \citenamefont {Savin},\ and\ \citenamefont {Pekola}}]{Giazotto2006review}%
  \BibitemOpen
  \bibfield  {author} {\bibinfo {author} {\bibfnamefont {F.}~\bibnamefont {Giazotto}}, \bibinfo {author} {\bibfnamefont {T.~T.}\ \bibnamefont {Heikkil\"a}}, \bibinfo {author} {\bibfnamefont {A.}~\bibnamefont {Luukanen}}, \bibinfo {author} {\bibfnamefont {A.~M.}\ \bibnamefont {Savin}},\ and\ \bibinfo {author} {\bibfnamefont {J.~P.}\ \bibnamefont {Pekola}},\ }\bibfield  {title} {\bibinfo {title} {Opportunities for mesoscopics in thermometry and refrigeration: Physics and applications},\ }\href {https://doi.org/10.1103/RevModPhys.78.217} {\bibfield  {journal} {\bibinfo  {journal} {Rev. Mod. Phys.}\ }\textbf {\bibinfo {volume} {78}},\ \bibinfo {pages} {217} (\bibinfo {year} {2006}{\natexlab{a}})}\BibitemShut {NoStop}%
\bibitem [{\citenamefont {Kawabata}\ \emph {et~al.}(2013)\citenamefont {Kawabata}, \citenamefont {Ozaeta}, \citenamefont {Vasenko}, \citenamefont {Hekking},\ and\ \citenamefont {Sebasti{\'a}n~Bergeret}}]{Kawabata2013}%
  \BibitemOpen
  \bibfield  {author} {\bibinfo {author} {\bibfnamefont {S.}~\bibnamefont {Kawabata}}, \bibinfo {author} {\bibfnamefont {A.}~\bibnamefont {Ozaeta}}, \bibinfo {author} {\bibfnamefont {A.~S.}\ \bibnamefont {Vasenko}}, \bibinfo {author} {\bibfnamefont {F.~W.~J.}\ \bibnamefont {Hekking}},\ and\ \bibinfo {author} {\bibfnamefont {F.}~\bibnamefont {Sebasti{\'a}n~Bergeret}},\ }\bibfield  {title} {\bibinfo {title} {Efficient electron refrigeration using superconductor/spin-filter devices},\ }\href {https://doi.org/10.1063/1.4813599} {\bibfield  {journal} {\bibinfo  {journal} {Applied Physics Letters}\ }\textbf {\bibinfo {volume} {103}},\ \bibinfo {pages} {032602} (\bibinfo {year} {2013})}\BibitemShut {NoStop}%
\bibitem [{\citenamefont {Huertas-Hernando}\ \emph {et~al.}(2002)\citenamefont {Huertas-Hernando}, \citenamefont {Nazarov},\ and\ \citenamefont {Belzig}}]{Huertas-Hernando2002}%
  \BibitemOpen
  \bibfield  {author} {\bibinfo {author} {\bibfnamefont {D.}~\bibnamefont {Huertas-Hernando}}, \bibinfo {author} {\bibfnamefont {Y.~V.}\ \bibnamefont {Nazarov}},\ and\ \bibinfo {author} {\bibfnamefont {W.}~\bibnamefont {Belzig}},\ }\bibfield  {title} {\bibinfo {title} {Absolute spin-valve effect with superconducting proximity structures},\ }\href {https://doi.org/10.1103/PhysRevLett.88.047003} {\bibfield  {journal} {\bibinfo  {journal} {Phys. Rev. Lett.}\ }\textbf {\bibinfo {volume} {88}},\ \bibinfo {pages} {047003} (\bibinfo {year} {2002})}\BibitemShut {NoStop}%
\bibitem [{\citenamefont {Giazotto}\ \emph {et~al.}(2006{\natexlab{b}})\citenamefont {Giazotto}, \citenamefont {Taddei}, \citenamefont {Fazio},\ and\ \citenamefont {Beltram}}]{Giazotto2006}%
  \BibitemOpen
  \bibfield  {author} {\bibinfo {author} {\bibfnamefont {F.}~\bibnamefont {Giazotto}}, \bibinfo {author} {\bibfnamefont {F.}~\bibnamefont {Taddei}}, \bibinfo {author} {\bibfnamefont {R.}~\bibnamefont {Fazio}},\ and\ \bibinfo {author} {\bibfnamefont {F.}~\bibnamefont {Beltram}},\ }\bibfield  {title} {\bibinfo {title} {Huge nonequilibrium magnetoresistance in hybrid superconducting spin valves},\ }\href {https://doi.org/10.1063/1.2220001} {\bibfield  {journal} {\bibinfo  {journal} {Applied Physics Letters}\ }\textbf {\bibinfo {volume} {89}},\ \bibinfo {pages} {022505} (\bibinfo {year} {2006}{\natexlab{b}})}\BibitemShut {NoStop}%
\bibitem [{\citenamefont {Giazotto}\ and\ \citenamefont {Taddei}(2008)}]{Giazotto2008}%
  \BibitemOpen
  \bibfield  {author} {\bibinfo {author} {\bibfnamefont {F.}~\bibnamefont {Giazotto}}\ and\ \bibinfo {author} {\bibfnamefont {F.}~\bibnamefont {Taddei}},\ }\bibfield  {title} {\bibinfo {title} {Superconductors as spin sources for spintronics},\ }\href {https://doi.org/10.1103/PhysRevB.77.132501} {\bibfield  {journal} {\bibinfo  {journal} {Phys. Rev. B}\ }\textbf {\bibinfo {volume} {77}},\ \bibinfo {pages} {132501} (\bibinfo {year} {2008})}\BibitemShut {NoStop}%
\bibitem [{\citenamefont {Giazotto}\ and\ \citenamefont {Bergeret}(2013)}]{Giazotto2013}%
  \BibitemOpen
  \bibfield  {author} {\bibinfo {author} {\bibfnamefont {F.}~\bibnamefont {Giazotto}}\ and\ \bibinfo {author} {\bibfnamefont {F.~S.}\ \bibnamefont {Bergeret}},\ }\bibfield  {title} {\bibinfo {title} {Quantum interference hybrid spin-current injector},\ }\href {https://doi.org/10.1063/1.4802953} {\bibfield  {journal} {\bibinfo  {journal} {Applied Physics Letters}\ }\textbf {\bibinfo {volume} {102}},\ \bibinfo {pages} {162406} (\bibinfo {year} {2013})}\BibitemShut {NoStop}%
\bibitem [{\citenamefont {Strambini}\ \emph {et~al.}(2017)\citenamefont {Strambini}, \citenamefont {Golovach}, \citenamefont {De~Simoni}, \citenamefont {Moodera}, \citenamefont {Bergeret},\ and\ \citenamefont {Giazotto}}]{Strambini2017}%
  \BibitemOpen
  \bibfield  {author} {\bibinfo {author} {\bibfnamefont {E.}~\bibnamefont {Strambini}}, \bibinfo {author} {\bibfnamefont {V.~N.}\ \bibnamefont {Golovach}}, \bibinfo {author} {\bibfnamefont {G.}~\bibnamefont {De~Simoni}}, \bibinfo {author} {\bibfnamefont {J.~S.}\ \bibnamefont {Moodera}}, \bibinfo {author} {\bibfnamefont {F.~S.}\ \bibnamefont {Bergeret}},\ and\ \bibinfo {author} {\bibfnamefont {F.}~\bibnamefont {Giazotto}},\ }\bibfield  {title} {\bibinfo {title} {Revealing the magnetic proximity effect in eus/al bilayers through superconducting tunneling spectroscopy},\ }\href {https://doi.org/10.1103/PhysRevMaterials.1.054402} {\bibfield  {journal} {\bibinfo  {journal} {Phys. Rev. Mater.}\ }\textbf {\bibinfo {volume} {1}},\ \bibinfo {pages} {054402} (\bibinfo {year} {2017})}\BibitemShut {NoStop}%
\bibitem [{\citenamefont {Giazotto}\ \emph {et~al.}(2015{\natexlab{b}})\citenamefont {Giazotto}, \citenamefont {Solinas}, \citenamefont {Braggio},\ and\ \citenamefont {Bergeret}}]{Giazotto2015_2}%
  \BibitemOpen
  \bibfield  {author} {\bibinfo {author} {\bibfnamefont {F.}~\bibnamefont {Giazotto}}, \bibinfo {author} {\bibfnamefont {P.}~\bibnamefont {Solinas}}, \bibinfo {author} {\bibfnamefont {A.}~\bibnamefont {Braggio}},\ and\ \bibinfo {author} {\bibfnamefont {F.~S.}\ \bibnamefont {Bergeret}},\ }\bibfield  {title} {\bibinfo {title} {Ferromagnetic-insulator-based superconducting junctions as sensitive electron thermometers},\ }\href {https://doi.org/10.1103/PhysRevApplied.4.044016} {\bibfield  {journal} {\bibinfo  {journal} {Phys. Rev. Appl.}\ }\textbf {\bibinfo {volume} {4}},\ \bibinfo {pages} {044016} (\bibinfo {year} {2015}{\natexlab{b}})}\BibitemShut {NoStop}%
\bibitem [{\citenamefont {Heikkil\"a}\ \emph {et~al.}(2018)\citenamefont {Heikkil\"a}, \citenamefont {Ojaj\"arvi}, \citenamefont {Maasilta}, \citenamefont {Strambini}, \citenamefont {Giazotto},\ and\ \citenamefont {Bergeret}}]{Heikkila2018}%
  \BibitemOpen
  \bibfield  {author} {\bibinfo {author} {\bibfnamefont {T.~T.}\ \bibnamefont {Heikkil\"a}}, \bibinfo {author} {\bibfnamefont {R.}~\bibnamefont {Ojaj\"arvi}}, \bibinfo {author} {\bibfnamefont {I.~J.}\ \bibnamefont {Maasilta}}, \bibinfo {author} {\bibfnamefont {E.}~\bibnamefont {Strambini}}, \bibinfo {author} {\bibfnamefont {F.}~\bibnamefont {Giazotto}},\ and\ \bibinfo {author} {\bibfnamefont {F.~S.}\ \bibnamefont {Bergeret}},\ }\bibfield  {title} {\bibinfo {title} {Thermoelectric radiation detector based on superconductor-ferromagnet systems},\ }\href {https://doi.org/10.1103/PhysRevApplied.10.034053} {\bibfield  {journal} {\bibinfo  {journal} {Phys. Rev. Appl.}\ }\textbf {\bibinfo {volume} {10}},\ \bibinfo {pages} {034053} (\bibinfo {year} {2018})}\BibitemShut {NoStop}%
\bibitem [{\citenamefont {Geng}\ \emph {et~al.}(2023)\citenamefont {Geng}, \citenamefont {Hijano}, \citenamefont {Ilić}, \citenamefont {Ilyn}, \citenamefont {Maasilta}, \citenamefont {Monfardini}, \citenamefont {Spies}, \citenamefont {Strambini}, \citenamefont {Virtanen}, \citenamefont {Calvo}, \citenamefont {González-Orellána}, \citenamefont {Helenius}, \citenamefont {Khorshidian}, \citenamefont {de~Araujo}, \citenamefont {Levy-Bertrand}, \citenamefont {Rogero}, \citenamefont {Giazotto}, \citenamefont {Bergeret},\ and\ \citenamefont {Heikkilä}}]{Geng2023}%
  \BibitemOpen
  \bibfield  {author} {\bibinfo {author} {\bibfnamefont {Z.}~\bibnamefont {Geng}}, \bibinfo {author} {\bibfnamefont {A.}~\bibnamefont {Hijano}}, \bibinfo {author} {\bibfnamefont {S.}~\bibnamefont {Ilić}}, \bibinfo {author} {\bibfnamefont {M.}~\bibnamefont {Ilyn}}, \bibinfo {author} {\bibfnamefont {I.}~\bibnamefont {Maasilta}}, \bibinfo {author} {\bibfnamefont {A.}~\bibnamefont {Monfardini}}, \bibinfo {author} {\bibfnamefont {M.}~\bibnamefont {Spies}}, \bibinfo {author} {\bibfnamefont {E.}~\bibnamefont {Strambini}}, \bibinfo {author} {\bibfnamefont {P.}~\bibnamefont {Virtanen}}, \bibinfo {author} {\bibfnamefont {M.}~\bibnamefont {Calvo}}, \bibinfo {author} {\bibfnamefont {C.}~\bibnamefont {González-Orellána}}, \bibinfo {author} {\bibfnamefont {A.~P.}\ \bibnamefont {Helenius}}, \bibinfo {author} {\bibfnamefont {S.}~\bibnamefont {Khorshidian}}, \bibinfo {author} {\bibfnamefont {C.~I.~L.}\ \bibnamefont {de~Araujo}}, \bibinfo {author} {\bibfnamefont {F.}~\bibnamefont {Levy-Bertrand}}, \bibinfo {author}
  {\bibfnamefont {C.}~\bibnamefont {Rogero}}, \bibinfo {author} {\bibfnamefont {F.}~\bibnamefont {Giazotto}}, \bibinfo {author} {\bibfnamefont {F.~S.}\ \bibnamefont {Bergeret}},\ and\ \bibinfo {author} {\bibfnamefont {T.~T.}\ \bibnamefont {Heikkilä}},\ }\bibfield  {title} {\bibinfo {title} {Superconductor-ferromagnet hybrids for non-reciprocal electronics and detectors},\ }\href {https://doi.org/10.1088/1361-6668/ad01e9} {\bibfield  {journal} {\bibinfo  {journal} {Superconductor Science and Technology}\ }\textbf {\bibinfo {volume} {36}},\ \bibinfo {pages} {123001} (\bibinfo {year} {2023})}\BibitemShut {NoStop}%
\bibitem [{\citenamefont {DE~GENNES}(1964)}]{deGennes1964}%
  \BibitemOpen
  \bibfield  {author} {\bibinfo {author} {\bibfnamefont {P.~G.}\ \bibnamefont {DE~GENNES}},\ }\bibfield  {title} {\bibinfo {title} {Boundary effects in superconductors},\ }\href {https://doi.org/10.1103/RevModPhys.36.225} {\bibfield  {journal} {\bibinfo  {journal} {Rev. Mod. Phys.}\ }\textbf {\bibinfo {volume} {36}},\ \bibinfo {pages} {225} (\bibinfo {year} {1964})}\BibitemShut {NoStop}%
\bibitem [{\citenamefont {Bergeret}\ \emph {et~al.}(2001)\citenamefont {Bergeret}, \citenamefont {Volkov},\ and\ \citenamefont {Efetov}}]{Bergeret2001}%
  \BibitemOpen
  \bibfield  {author} {\bibinfo {author} {\bibfnamefont {F.~S.}\ \bibnamefont {Bergeret}}, \bibinfo {author} {\bibfnamefont {A.~F.}\ \bibnamefont {Volkov}},\ and\ \bibinfo {author} {\bibfnamefont {K.~B.}\ \bibnamefont {Efetov}},\ }\bibfield  {title} {\bibinfo {title} {Enhancement of the josephson current by an exchange field in superconductor-ferromagnet structures},\ }\href {https://doi.org/10.1103/PhysRevLett.86.3140} {\bibfield  {journal} {\bibinfo  {journal} {Phys. Rev. Lett.}\ }\textbf {\bibinfo {volume} {86}},\ \bibinfo {pages} {3140} (\bibinfo {year} {2001})}\BibitemShut {NoStop}%
\bibitem [{\citenamefont {Tokuyasu}\ \emph {et~al.}(1988)\citenamefont {Tokuyasu}, \citenamefont {Sauls},\ and\ \citenamefont {Rainer}}]{Tokuyasu1988}%
  \BibitemOpen
  \bibfield  {author} {\bibinfo {author} {\bibfnamefont {T.}~\bibnamefont {Tokuyasu}}, \bibinfo {author} {\bibfnamefont {J.~A.}\ \bibnamefont {Sauls}},\ and\ \bibinfo {author} {\bibfnamefont {D.}~\bibnamefont {Rainer}},\ }\bibfield  {title} {\bibinfo {title} {Proximity effect of a ferromagnetic insulator in contact with a superconductor},\ }\href {https://doi.org/10.1103/PhysRevB.38.8823} {\bibfield  {journal} {\bibinfo  {journal} {Phys. Rev. B}\ }\textbf {\bibinfo {volume} {38}},\ \bibinfo {pages} {8823} (\bibinfo {year} {1988})}\BibitemShut {NoStop}%
\bibitem [{\citenamefont {Cottet}\ \emph {et~al.}(2009)\citenamefont {Cottet}, \citenamefont {Huertas-Hernando}, \citenamefont {Belzig},\ and\ \citenamefont {Nazarov}}]{Cottet2009}%
  \BibitemOpen
  \bibfield  {author} {\bibinfo {author} {\bibfnamefont {A.}~\bibnamefont {Cottet}}, \bibinfo {author} {\bibfnamefont {D.}~\bibnamefont {Huertas-Hernando}}, \bibinfo {author} {\bibfnamefont {W.}~\bibnamefont {Belzig}},\ and\ \bibinfo {author} {\bibfnamefont {Y.~V.}\ \bibnamefont {Nazarov}},\ }\bibfield  {title} {\bibinfo {title} {Spin-dependent boundary conditions for isotropic superconducting green's functions},\ }\href {https://doi.org/10.1103/PhysRevB.80.184511} {\bibfield  {journal} {\bibinfo  {journal} {Phys. Rev. B}\ }\textbf {\bibinfo {volume} {80}},\ \bibinfo {pages} {184511} (\bibinfo {year} {2009})}\BibitemShut {NoStop}%
\bibitem [{\citenamefont {Eschrig}\ \emph {et~al.}(2015)\citenamefont {Eschrig}, \citenamefont {Cottet}, \citenamefont {Belzig},\ and\ \citenamefont {Linder}}]{Eschrig2015_bc}%
  \BibitemOpen
  \bibfield  {author} {\bibinfo {author} {\bibfnamefont {M.}~\bibnamefont {Eschrig}}, \bibinfo {author} {\bibfnamefont {A.}~\bibnamefont {Cottet}}, \bibinfo {author} {\bibfnamefont {W.}~\bibnamefont {Belzig}},\ and\ \bibinfo {author} {\bibfnamefont {J.}~\bibnamefont {Linder}},\ }\bibfield  {title} {\bibinfo {title} {General boundary conditions for quasiclassical theory of superconductivity in the diffusive limit: application to strongly spin-polarized systems},\ }\href {https://doi.org/10.1088/1367-2630/17/8/083037} {\bibfield  {journal} {\bibinfo  {journal} {New Journal of Physics}\ }\textbf {\bibinfo {volume} {17}},\ \bibinfo {pages} {083037} (\bibinfo {year} {2015})}\BibitemShut {NoStop}%
\bibitem [{\citenamefont {Aikebaier}\ \emph {et~al.}(2022)\citenamefont {Aikebaier}, \citenamefont {Heikkil\"a},\ and\ \citenamefont {Lado}}]{Aikebaier2022}%
  \BibitemOpen
  \bibfield  {author} {\bibinfo {author} {\bibfnamefont {F.}~\bibnamefont {Aikebaier}}, \bibinfo {author} {\bibfnamefont {T.~T.}\ \bibnamefont {Heikkil\"a}},\ and\ \bibinfo {author} {\bibfnamefont {J.~L.}\ \bibnamefont {Lado}},\ }\bibfield  {title} {\bibinfo {title} {Controlling magnetism through ising superconductivity in magnetic van der waals heterostructures},\ }\href {https://doi.org/10.1103/PhysRevB.105.054506} {\bibfield  {journal} {\bibinfo  {journal} {Phys. Rev. B}\ }\textbf {\bibinfo {volume} {105}},\ \bibinfo {pages} {054506} (\bibinfo {year} {2022})}\BibitemShut {NoStop}%
\bibitem [{\citenamefont {Kang}\ \emph {et~al.}(2021)\citenamefont {Kang}, \citenamefont {Jiang}, \citenamefont {Berger}, \citenamefont {Watanabe}, \citenamefont {Taniguchi}, \citenamefont {Forro}, \citenamefont {Shan},\ and\ \citenamefont {Mak}}]{Kang2021}%
  \BibitemOpen
  \bibfield  {author} {\bibinfo {author} {\bibfnamefont {K.}~\bibnamefont {Kang}}, \bibinfo {author} {\bibfnamefont {S.}~\bibnamefont {Jiang}}, \bibinfo {author} {\bibfnamefont {H.}~\bibnamefont {Berger}}, \bibinfo {author} {\bibfnamefont {K.}~\bibnamefont {Watanabe}}, \bibinfo {author} {\bibfnamefont {T.}~\bibnamefont {Taniguchi}}, \bibinfo {author} {\bibfnamefont {L.}~\bibnamefont {Forro}}, \bibinfo {author} {\bibfnamefont {J.}~\bibnamefont {Shan}},\ and\ \bibinfo {author} {\bibfnamefont {K.~F.}\ \bibnamefont {Mak}},\ }\href@noop {} {\bibinfo {title} {Giant anisotropic magnetoresistance in ising superconductor-magnetic insulator tunnel junctions}} (\bibinfo {year} {2021}),\ \Eprint {https://arxiv.org/abs/2101.01327} {arXiv:2101.01327 [cond-mat.supr-con]} \BibitemShut {NoStop}%
\bibitem [{\citenamefont {Wickramaratne}\ \emph {et~al.}(2021)\citenamefont {Wickramaratne}, \citenamefont {Haim}, \citenamefont {Khodas},\ and\ \citenamefont {Mazin}}]{Wickramaratne2021}%
  \BibitemOpen
  \bibfield  {author} {\bibinfo {author} {\bibfnamefont {D.}~\bibnamefont {Wickramaratne}}, \bibinfo {author} {\bibfnamefont {M.}~\bibnamefont {Haim}}, \bibinfo {author} {\bibfnamefont {M.}~\bibnamefont {Khodas}},\ and\ \bibinfo {author} {\bibfnamefont {I.~I.}\ \bibnamefont {Mazin}},\ }\bibfield  {title} {\bibinfo {title} {Magnetism-driven unconventional effects in ising superconductors: Role of proximity, tunneling, and nematicity},\ }\href {https://doi.org/10.1103/PhysRevB.104.L060501} {\bibfield  {journal} {\bibinfo  {journal} {Phys. Rev. B}\ }\textbf {\bibinfo {volume} {104}},\ \bibinfo {pages} {L060501} (\bibinfo {year} {2021})}\BibitemShut {NoStop}%
\bibitem [{\citenamefont {Jo}\ \emph {et~al.}(2023)\citenamefont {Jo}, \citenamefont {Peisen}, \citenamefont {Yang}, \citenamefont {Ma{\~{n}}as-Valero}, \citenamefont {Baldov{\'i}}, \citenamefont {Lu}, \citenamefont {Coronado}, \citenamefont {Casanova}, \citenamefont {Bergeret}, \citenamefont {Gobbi},\ and\ \citenamefont {Hueso}}]{Jo2023}%
  \BibitemOpen
  \bibfield  {author} {\bibinfo {author} {\bibfnamefont {J.}~\bibnamefont {Jo}}, \bibinfo {author} {\bibfnamefont {Y.}~\bibnamefont {Peisen}}, \bibinfo {author} {\bibfnamefont {H.}~\bibnamefont {Yang}}, \bibinfo {author} {\bibfnamefont {S.}~\bibnamefont {Ma{\~{n}}as-Valero}}, \bibinfo {author} {\bibfnamefont {J.~J.}\ \bibnamefont {Baldov{\'i}}}, \bibinfo {author} {\bibfnamefont {Y.}~\bibnamefont {Lu}}, \bibinfo {author} {\bibfnamefont {E.}~\bibnamefont {Coronado}}, \bibinfo {author} {\bibfnamefont {F.}~\bibnamefont {Casanova}}, \bibinfo {author} {\bibfnamefont {F.~S.}\ \bibnamefont {Bergeret}}, \bibinfo {author} {\bibfnamefont {M.}~\bibnamefont {Gobbi}},\ and\ \bibinfo {author} {\bibfnamefont {L.~E.}\ \bibnamefont {Hueso}},\ }\bibfield  {title} {\bibinfo {title} {Local control of superconductivity in a nbse2/crsbr van der waals heterostructure},\ }\href {https://doi.org/10.1038/s41467-023-43111-7} {\bibfield  {journal} {\bibinfo  {journal} {Nature Communications}\ }\textbf {\bibinfo {volume} {14}},\ \bibinfo
  {pages} {7253} (\bibinfo {year} {2023})}\BibitemShut {NoStop}%
\bibitem [{\citenamefont {Jiang}\ \emph {et~al.}(2020)\citenamefont {Jiang}, \citenamefont {Yuan}, \citenamefont {Wu}, \citenamefont {Wei}, \citenamefont {Mu}, \citenamefont {An},\ and\ \citenamefont {Li}}]{Jiang2020}%
  \BibitemOpen
  \bibfield  {author} {\bibinfo {author} {\bibfnamefont {D.}~\bibnamefont {Jiang}}, \bibinfo {author} {\bibfnamefont {T.}~\bibnamefont {Yuan}}, \bibinfo {author} {\bibfnamefont {Y.}~\bibnamefont {Wu}}, \bibinfo {author} {\bibfnamefont {X.}~\bibnamefont {Wei}}, \bibinfo {author} {\bibfnamefont {G.}~\bibnamefont {Mu}}, \bibinfo {author} {\bibfnamefont {Z.}~\bibnamefont {An}},\ and\ \bibinfo {author} {\bibfnamefont {W.}~\bibnamefont {Li}},\ }\bibfield  {title} {\bibinfo {title} {Strong in-plane magnetic field-induced reemergent superconductivity in the van der waals heterointerface of nbse2 and crcl3},\ }\href {https://doi.org/10.1021/acsami.0c15203} {\bibfield  {journal} {\bibinfo  {journal} {ACS Applied Materials {\&} Interfaces}\ }\textbf {\bibinfo {volume} {12}},\ \bibinfo {pages} {49252} (\bibinfo {year} {2020})}\BibitemShut {NoStop}%
\bibitem [{\citenamefont {Kezilebieke}\ \emph {et~al.}(2020)\citenamefont {Kezilebieke}, \citenamefont {Huda}, \citenamefont {Va{\v{n}}o}, \citenamefont {Aapro}, \citenamefont {Ganguli}, \citenamefont {Silveira}, \citenamefont {G{\l}odzik}, \citenamefont {Foster}, \citenamefont {Ojanen},\ and\ \citenamefont {Liljeroth}}]{Kezilebieke2020}%
  \BibitemOpen
  \bibfield  {author} {\bibinfo {author} {\bibfnamefont {S.}~\bibnamefont {Kezilebieke}}, \bibinfo {author} {\bibfnamefont {M.~N.}\ \bibnamefont {Huda}}, \bibinfo {author} {\bibfnamefont {V.}~\bibnamefont {Va{\v{n}}o}}, \bibinfo {author} {\bibfnamefont {M.}~\bibnamefont {Aapro}}, \bibinfo {author} {\bibfnamefont {S.~C.}\ \bibnamefont {Ganguli}}, \bibinfo {author} {\bibfnamefont {O.~J.}\ \bibnamefont {Silveira}}, \bibinfo {author} {\bibfnamefont {S.}~\bibnamefont {G{\l}odzik}}, \bibinfo {author} {\bibfnamefont {A.~S.}\ \bibnamefont {Foster}}, \bibinfo {author} {\bibfnamefont {T.}~\bibnamefont {Ojanen}},\ and\ \bibinfo {author} {\bibfnamefont {P.}~\bibnamefont {Liljeroth}},\ }\bibfield  {title} {\bibinfo {title} {Topological superconductivity in a van der waals heterostructure},\ }\href {https://doi.org/10.1038/s41586-020-2989-y} {\bibfield  {journal} {\bibinfo  {journal} {Nature}\ }\textbf {\bibinfo {volume} {588}},\ \bibinfo {pages} {424} (\bibinfo {year} {2020})}\BibitemShut {NoStop}%
\bibitem [{\citenamefont {Ai}\ \emph {et~al.}(2021)\citenamefont {Ai}, \citenamefont {Zhang}, \citenamefont {Yang}, \citenamefont {Xie}, \citenamefont {Yang}, \citenamefont {Jia}, \citenamefont {Zhang}, \citenamefont {Liu}, \citenamefont {Li}, \citenamefont {Leng}, \citenamefont {Cao}, \citenamefont {Sun}, \citenamefont {Zhang}, \citenamefont {Kou}, \citenamefont {Han}, \citenamefont {Xiu},\ and\ \citenamefont {Dong}}]{Ai2021}%
  \BibitemOpen
  \bibfield  {author} {\bibinfo {author} {\bibfnamefont {L.}~\bibnamefont {Ai}}, \bibinfo {author} {\bibfnamefont {E.}~\bibnamefont {Zhang}}, \bibinfo {author} {\bibfnamefont {J.}~\bibnamefont {Yang}}, \bibinfo {author} {\bibfnamefont {X.}~\bibnamefont {Xie}}, \bibinfo {author} {\bibfnamefont {Y.}~\bibnamefont {Yang}}, \bibinfo {author} {\bibfnamefont {Z.}~\bibnamefont {Jia}}, \bibinfo {author} {\bibfnamefont {Y.}~\bibnamefont {Zhang}}, \bibinfo {author} {\bibfnamefont {S.}~\bibnamefont {Liu}}, \bibinfo {author} {\bibfnamefont {Z.}~\bibnamefont {Li}}, \bibinfo {author} {\bibfnamefont {P.}~\bibnamefont {Leng}}, \bibinfo {author} {\bibfnamefont {X.}~\bibnamefont {Cao}}, \bibinfo {author} {\bibfnamefont {X.}~\bibnamefont {Sun}}, \bibinfo {author} {\bibfnamefont {T.}~\bibnamefont {Zhang}}, \bibinfo {author} {\bibfnamefont {X.}~\bibnamefont {Kou}}, \bibinfo {author} {\bibfnamefont {Z.}~\bibnamefont {Han}}, \bibinfo {author} {\bibfnamefont {F.}~\bibnamefont {Xiu}},\ and\ \bibinfo {author} {\bibfnamefont
  {S.}~\bibnamefont {Dong}},\ }\bibfield  {title} {\bibinfo {title} {Van der waals ferromagnetic josephson junctions},\ }\href {https://doi.org/10.1038/s41467-021-26946-w} {\bibfield  {journal} {\bibinfo  {journal} {Nature Communications}\ }\textbf {\bibinfo {volume} {12}},\ \bibinfo {pages} {6580} (\bibinfo {year} {2021})}\BibitemShut {NoStop}%
\bibitem [{\citenamefont {Idzuchi}\ \emph {et~al.}(2021)\citenamefont {Idzuchi}, \citenamefont {Pientka}, \citenamefont {Huang}, \citenamefont {Harada}, \citenamefont {G{\"u}l}, \citenamefont {Shin}, \citenamefont {Nguyen}, \citenamefont {Jo}, \citenamefont {Shindo}, \citenamefont {Cava}, \citenamefont {Canfield},\ and\ \citenamefont {Kim}}]{Idzuchi2021}%
  \BibitemOpen
  \bibfield  {author} {\bibinfo {author} {\bibfnamefont {H.}~\bibnamefont {Idzuchi}}, \bibinfo {author} {\bibfnamefont {F.}~\bibnamefont {Pientka}}, \bibinfo {author} {\bibfnamefont {K.-F.}\ \bibnamefont {Huang}}, \bibinfo {author} {\bibfnamefont {K.}~\bibnamefont {Harada}}, \bibinfo {author} {\bibfnamefont {{\"O}.}~\bibnamefont {G{\"u}l}}, \bibinfo {author} {\bibfnamefont {Y.~J.}\ \bibnamefont {Shin}}, \bibinfo {author} {\bibfnamefont {L.~T.}\ \bibnamefont {Nguyen}}, \bibinfo {author} {\bibfnamefont {N.~H.}\ \bibnamefont {Jo}}, \bibinfo {author} {\bibfnamefont {D.}~\bibnamefont {Shindo}}, \bibinfo {author} {\bibfnamefont {R.~J.}\ \bibnamefont {Cava}}, \bibinfo {author} {\bibfnamefont {P.~C.}\ \bibnamefont {Canfield}},\ and\ \bibinfo {author} {\bibfnamefont {P.}~\bibnamefont {Kim}},\ }\bibfield  {title} {\bibinfo {title} {Unconventional supercurrent phase in ising superconductor josephson junction with atomically thin magnetic insulator},\ }\href {https://doi.org/10.1038/s41467-021-25608-1} {\bibfield
  {journal} {\bibinfo  {journal} {Nature Communications}\ }\textbf {\bibinfo {volume} {12}},\ \bibinfo {pages} {5332} (\bibinfo {year} {2021})}\BibitemShut {NoStop}%
\bibitem [{\citenamefont {Bobkov}\ \emph {et~al.}(2024{\natexlab{a}})\citenamefont {Bobkov}, \citenamefont {Bokai}, \citenamefont {Otrokov}, \citenamefont {Bobkov},\ and\ \citenamefont {Bobkova}}]{Bobkov2024_vdW}%
  \BibitemOpen
  \bibfield  {author} {\bibinfo {author} {\bibfnamefont {G.~A.}\ \bibnamefont {Bobkov}}, \bibinfo {author} {\bibfnamefont {K.~A.}\ \bibnamefont {Bokai}}, \bibinfo {author} {\bibfnamefont {M.~M.}\ \bibnamefont {Otrokov}}, \bibinfo {author} {\bibfnamefont {A.~M.}\ \bibnamefont {Bobkov}},\ and\ \bibinfo {author} {\bibfnamefont {I.~V.}\ \bibnamefont {Bobkova}},\ }\bibfield  {title} {\bibinfo {title} {Gate-controlled proximity effect in superconductor/ferromagnet van der waals heterostructures},\ }\href {https://doi.org/10.1103/PhysRevMaterials.8.104801} {\bibfield  {journal} {\bibinfo  {journal} {Phys. Rev. Mater.}\ }\textbf {\bibinfo {volume} {8}},\ \bibinfo {pages} {104801} (\bibinfo {year} {2024}{\natexlab{a}})}\BibitemShut {NoStop}%
\bibitem [{\citenamefont {Bobkov}\ \emph {et~al.}(2024{\natexlab{b}})\citenamefont {Bobkov}, \citenamefont {Bobkov},\ and\ \citenamefont {Bobkova}}]{Bobkov2024_spin}%
  \BibitemOpen
  \bibfield  {author} {\bibinfo {author} {\bibfnamefont {G.~A.}\ \bibnamefont {Bobkov}}, \bibinfo {author} {\bibfnamefont {A.~M.}\ \bibnamefont {Bobkov}},\ and\ \bibinfo {author} {\bibfnamefont {I.~V.}\ \bibnamefont {Bobkova}},\ }\bibfield  {title} {\bibinfo {title} {Spin supercurrent in superconductor/ferromagnet van der waals heterostructures},\ }\href {https://doi.org/10.1103/PhysRevB.110.104506} {\bibfield  {journal} {\bibinfo  {journal} {Phys. Rev. B}\ }\textbf {\bibinfo {volume} {110}},\ \bibinfo {pages} {104506} (\bibinfo {year} {2024}{\natexlab{b}})}\BibitemShut {NoStop}%
\bibitem [{\citenamefont {Wang}\ \emph {et~al.}(2021)\citenamefont {Wang}, \citenamefont {Xu},\ and\ \citenamefont {Duan}}]{Wang2021}%
  \BibitemOpen
  \bibfield  {author} {\bibinfo {author} {\bibfnamefont {C.}~\bibnamefont {Wang}}, \bibinfo {author} {\bibfnamefont {Y.}~\bibnamefont {Xu}},\ and\ \bibinfo {author} {\bibfnamefont {W.}~\bibnamefont {Duan}},\ }\bibfield  {title} {\bibinfo {title} {Ising superconductivity and its hidden variants},\ }\href {https://doi.org/10.1021/accountsmr.1c00068} {\bibfield  {journal} {\bibinfo  {journal} {Accounts of Materials Research}\ }\textbf {\bibinfo {volume} {2}},\ \bibinfo {pages} {526} (\bibinfo {year} {2021})}\BibitemShut {NoStop}%
\bibitem [{\citenamefont {Kaur}\ \emph {et~al.}(2005)\citenamefont {Kaur}, \citenamefont {Agterberg},\ and\ \citenamefont {Sigrist}}]{Kaur2005}%
  \BibitemOpen
  \bibfield  {author} {\bibinfo {author} {\bibfnamefont {R.~P.}\ \bibnamefont {Kaur}}, \bibinfo {author} {\bibfnamefont {D.~F.}\ \bibnamefont {Agterberg}},\ and\ \bibinfo {author} {\bibfnamefont {M.}~\bibnamefont {Sigrist}},\ }\bibfield  {title} {\bibinfo {title} {Helical vortex phase in the noncentrosymmetric ${\mathrm{cept}}_{3}\mathrm{Si}$},\ }\href {https://doi.org/10.1103/PhysRevLett.94.137002} {\bibfield  {journal} {\bibinfo  {journal} {Phys. Rev. Lett.}\ }\textbf {\bibinfo {volume} {94}},\ \bibinfo {pages} {137002} (\bibinfo {year} {2005})}\BibitemShut {NoStop}%
\bibitem [{\citenamefont {Akbari}\ and\ \citenamefont {Thalmeier}(2022)}]{Akbari2022}%
  \BibitemOpen
  \bibfield  {author} {\bibinfo {author} {\bibfnamefont {A.}~\bibnamefont {Akbari}}\ and\ \bibinfo {author} {\bibfnamefont {P.}~\bibnamefont {Thalmeier}},\ }\bibfield  {title} {\bibinfo {title} {Fermi surface segmentation in the helical state of a rashba superconductor},\ }\href {https://doi.org/10.1103/PhysRevResearch.4.023096} {\bibfield  {journal} {\bibinfo  {journal} {Phys. Rev. Res.}\ }\textbf {\bibinfo {volume} {4}},\ \bibinfo {pages} {023096} (\bibinfo {year} {2022})}\BibitemShut {NoStop}%
\bibitem [{\citenamefont {Zhang}\ and\ \citenamefont {Liu}(2022)}]{Zhang2022}%
  \BibitemOpen
  \bibfield  {author} {\bibinfo {author} {\bibfnamefont {X.}~\bibnamefont {Zhang}}\ and\ \bibinfo {author} {\bibfnamefont {F.}~\bibnamefont {Liu}},\ }\bibfield  {title} {\bibinfo {title} {Fulde-ferrell-larkin-ovchinnikov pairing induced by a weyl nodal line in an ising superconductor with a high critical field},\ }\href {https://doi.org/10.1103/PhysRevB.105.024505} {\bibfield  {journal} {\bibinfo  {journal} {Phys. Rev. B}\ }\textbf {\bibinfo {volume} {105}},\ \bibinfo {pages} {024505} (\bibinfo {year} {2022})}\BibitemShut {NoStop}%
\bibitem [{\citenamefont {Zhao}\ \emph {et~al.}(2023)\citenamefont {Zhao}, \citenamefont {Debbeler}, \citenamefont {K{\"u}hne}, \citenamefont {Fecher}, \citenamefont {Gross},\ and\ \citenamefont {Smet}}]{Zhao2023}%
  \BibitemOpen
  \bibfield  {author} {\bibinfo {author} {\bibfnamefont {D.}~\bibnamefont {Zhao}}, \bibinfo {author} {\bibfnamefont {L.}~\bibnamefont {Debbeler}}, \bibinfo {author} {\bibfnamefont {M.}~\bibnamefont {K{\"u}hne}}, \bibinfo {author} {\bibfnamefont {S.}~\bibnamefont {Fecher}}, \bibinfo {author} {\bibfnamefont {N.}~\bibnamefont {Gross}},\ and\ \bibinfo {author} {\bibfnamefont {J.}~\bibnamefont {Smet}},\ }\bibfield  {title} {\bibinfo {title} {Evidence of finite-momentum pairing in a centrosymmetric bilayer},\ }\href {https://doi.org/10.1038/s41567-023-02202-4} {\bibfield  {journal} {\bibinfo  {journal} {Nature Physics}\ }\textbf {\bibinfo {volume} {19}},\ \bibinfo {pages} {1599} (\bibinfo {year} {2023})}\BibitemShut {NoStop}%
\bibitem [{\citenamefont {Wan}\ \emph {et~al.}(2023)\citenamefont {Wan}, \citenamefont {Zheliuk}, \citenamefont {Yuan}, \citenamefont {Peng}, \citenamefont {Zhang}, \citenamefont {Liang}, \citenamefont {Zeitler}, \citenamefont {Wiedmann}, \citenamefont {Hussey}, \citenamefont {Palstra},\ and\ \citenamefont {Ye}}]{Wan2023}%
  \BibitemOpen
  \bibfield  {author} {\bibinfo {author} {\bibfnamefont {P.}~\bibnamefont {Wan}}, \bibinfo {author} {\bibfnamefont {O.}~\bibnamefont {Zheliuk}}, \bibinfo {author} {\bibfnamefont {N.~F.~Q.}\ \bibnamefont {Yuan}}, \bibinfo {author} {\bibfnamefont {X.}~\bibnamefont {Peng}}, \bibinfo {author} {\bibfnamefont {L.}~\bibnamefont {Zhang}}, \bibinfo {author} {\bibfnamefont {M.}~\bibnamefont {Liang}}, \bibinfo {author} {\bibfnamefont {U.}~\bibnamefont {Zeitler}}, \bibinfo {author} {\bibfnamefont {S.}~\bibnamefont {Wiedmann}}, \bibinfo {author} {\bibfnamefont {N.~E.}\ \bibnamefont {Hussey}}, \bibinfo {author} {\bibfnamefont {T.~T.~M.}\ \bibnamefont {Palstra}},\ and\ \bibinfo {author} {\bibfnamefont {J.}~\bibnamefont {Ye}},\ }\bibfield  {title} {\bibinfo {title} {Orbital fulde--ferrell--larkin--ovchinnikov state in an ising superconductor},\ }\href {https://doi.org/10.1038/s41586-023-05967-z} {\bibfield  {journal} {\bibinfo  {journal} {Nature}\ }\textbf {\bibinfo {volume} {619}},\ \bibinfo {pages} {46} (\bibinfo {year}
  {2023})}\BibitemShut {NoStop}%
\bibitem [{\citenamefont {Ding}\ \emph {et~al.}(2024)\citenamefont {Ding}, \citenamefont {He}, \citenamefont {Zhang}, \citenamefont {Zuo}, \citenamefont {Gu}, \citenamefont {Cai}, \citenamefont {Zeng}, \citenamefont {Yan}, \citenamefont {Cai}, \citenamefont {Cao}, \citenamefont {Watanabe}, \citenamefont {Taniguchi}, \citenamefont {Dong}, \citenamefont {Zhang}, \citenamefont {Wu}, \citenamefont {Zhou}, \citenamefont {Wang}, \citenamefont {Chen}, \citenamefont {Ye}, \citenamefont {Liu},\ and\ \citenamefont {Li}}]{Ding2023}%
  \BibitemOpen
  \bibfield  {author} {\bibinfo {author} {\bibfnamefont {Y.}~\bibnamefont {Ding}}, \bibinfo {author} {\bibfnamefont {J.}~\bibnamefont {He}}, \bibinfo {author} {\bibfnamefont {S.}~\bibnamefont {Zhang}}, \bibinfo {author} {\bibfnamefont {H.}~\bibnamefont {Zuo}}, \bibinfo {author} {\bibfnamefont {P.}~\bibnamefont {Gu}}, \bibinfo {author} {\bibfnamefont {J.}~\bibnamefont {Cai}}, \bibinfo {author} {\bibfnamefont {X.}~\bibnamefont {Zeng}}, \bibinfo {author} {\bibfnamefont {P.}~\bibnamefont {Yan}}, \bibinfo {author} {\bibfnamefont {J.}~\bibnamefont {Cai}}, \bibinfo {author} {\bibfnamefont {K.}~\bibnamefont {Cao}}, \bibinfo {author} {\bibfnamefont {K.}~\bibnamefont {Watanabe}}, \bibinfo {author} {\bibfnamefont {T.}~\bibnamefont {Taniguchi}}, \bibinfo {author} {\bibfnamefont {P.}~\bibnamefont {Dong}}, \bibinfo {author} {\bibfnamefont {Y.}~\bibnamefont {Zhang}}, \bibinfo {author} {\bibfnamefont {Y.}~\bibnamefont {Wu}}, \bibinfo {author} {\bibfnamefont {X.}~\bibnamefont {Zhou}}, \bibinfo {author} {\bibfnamefont
  {J.}~\bibnamefont {Wang}}, \bibinfo {author} {\bibfnamefont {Y.}~\bibnamefont {Chen}}, \bibinfo {author} {\bibfnamefont {Y.}~\bibnamefont {Ye}}, \bibinfo {author} {\bibfnamefont {J.}~\bibnamefont {Liu}},\ and\ \bibinfo {author} {\bibfnamefont {J.}~\bibnamefont {Li}},\ }\bibfield  {title} {\bibinfo {title} {Constructing the fulde–ferrell–larkin–ovchinnikov state in a crocl/nbse2 van der waals heterostructure},\ }\href {https://doi.org/10.1021/acs.nanolett.4c03079} {\bibfield  {journal} {\bibinfo  {journal} {Nano Letters}\ }\textbf {\bibinfo {volume} {24}},\ \bibinfo {pages} {12814} (\bibinfo {year} {2024})},\ \bibinfo {note} {pMID: 39361493},\ \Eprint {https://arxiv.org/abs/https://doi.org/10.1021/acs.nanolett.4c03079} {https://doi.org/10.1021/acs.nanolett.4c03079} \BibitemShut {NoStop}%
\bibitem [{\citenamefont {Sarma}(1963)}]{Sarma1963}%
  \BibitemOpen
  \bibfield  {author} {\bibinfo {author} {\bibfnamefont {G.}~\bibnamefont {Sarma}},\ }\bibfield  {title} {\bibinfo {title} {On the influence of a uniform exchange field acting on the spins of the conduction electrons in a superconductor},\ }\href {https://www.sciencedirect.com/science/article/pii/0022369763900076} {\bibfield  {journal} {\bibinfo  {journal} {Journal of Physics and Chemistry of Solids}\ }\textbf {\bibinfo {volume} {24}},\ \bibinfo {pages} {1029} (\bibinfo {year} {1963})}\BibitemShut {NoStop}%
\bibitem [{\citenamefont {Xi}\ \emph {et~al.}(2016{\natexlab{a}})\citenamefont {Xi}, \citenamefont {Wang}, \citenamefont {Zhao}, \citenamefont {Park}, \citenamefont {Law}, \citenamefont {Berger}, \citenamefont {Forr{\'o}}, \citenamefont {Shan},\ and\ \citenamefont {Mak}}]{Xi2016_2}%
  \BibitemOpen
  \bibfield  {author} {\bibinfo {author} {\bibfnamefont {X.}~\bibnamefont {Xi}}, \bibinfo {author} {\bibfnamefont {Z.}~\bibnamefont {Wang}}, \bibinfo {author} {\bibfnamefont {W.}~\bibnamefont {Zhao}}, \bibinfo {author} {\bibfnamefont {J.-H.}\ \bibnamefont {Park}}, \bibinfo {author} {\bibfnamefont {K.~T.}\ \bibnamefont {Law}}, \bibinfo {author} {\bibfnamefont {H.}~\bibnamefont {Berger}}, \bibinfo {author} {\bibfnamefont {L.}~\bibnamefont {Forr{\'o}}}, \bibinfo {author} {\bibfnamefont {J.}~\bibnamefont {Shan}},\ and\ \bibinfo {author} {\bibfnamefont {K.~F.}\ \bibnamefont {Mak}},\ }\bibfield  {title} {\bibinfo {title} {Ising pairing in superconducting nbse2 atomic layers},\ }\href {https://doi.org/10.1038/nphys3538} {\bibfield  {journal} {\bibinfo  {journal} {Nature Physics}\ }\textbf {\bibinfo {volume} {12}},\ \bibinfo {pages} {139} (\bibinfo {year} {2016}{\natexlab{a}})}\BibitemShut {NoStop}%
\bibitem [{\citenamefont {de~la Barrera}\ \emph {et~al.}(2018)\citenamefont {de~la Barrera}, \citenamefont {Sinko}, \citenamefont {Gopalan}, \citenamefont {Sivadas}, \citenamefont {Seyler}, \citenamefont {Watanabe}, \citenamefont {Taniguchi}, \citenamefont {Tsen}, \citenamefont {Xu}, \citenamefont {Xiao},\ and\ \citenamefont {Hunt}}]{delaBarrera2018}%
  \BibitemOpen
  \bibfield  {author} {\bibinfo {author} {\bibfnamefont {S.~C.}\ \bibnamefont {de~la Barrera}}, \bibinfo {author} {\bibfnamefont {M.~R.}\ \bibnamefont {Sinko}}, \bibinfo {author} {\bibfnamefont {D.~P.}\ \bibnamefont {Gopalan}}, \bibinfo {author} {\bibfnamefont {N.}~\bibnamefont {Sivadas}}, \bibinfo {author} {\bibfnamefont {K.~L.}\ \bibnamefont {Seyler}}, \bibinfo {author} {\bibfnamefont {K.}~\bibnamefont {Watanabe}}, \bibinfo {author} {\bibfnamefont {T.}~\bibnamefont {Taniguchi}}, \bibinfo {author} {\bibfnamefont {A.~W.}\ \bibnamefont {Tsen}}, \bibinfo {author} {\bibfnamefont {X.}~\bibnamefont {Xu}}, \bibinfo {author} {\bibfnamefont {D.}~\bibnamefont {Xiao}},\ and\ \bibinfo {author} {\bibfnamefont {B.~M.}\ \bibnamefont {Hunt}},\ }\bibfield  {title} {\bibinfo {title} {Tuning ising superconductivity with layer and spin--orbit coupling in two-dimensional transition-metal dichalcogenides},\ }\href {https://doi.org/10.1038/s41467-018-03888-4} {\bibfield  {journal} {\bibinfo  {journal} {Nature Communications}\
  }\textbf {\bibinfo {volume} {9}},\ \bibinfo {pages} {1427} (\bibinfo {year} {2018})}\BibitemShut {NoStop}%
\bibitem [{\citenamefont {Dvir}\ \emph {et~al.}(2018)\citenamefont {Dvir}, \citenamefont {Massee}, \citenamefont {Attias}, \citenamefont {Khodas}, \citenamefont {Aprili}, \citenamefont {Quay},\ and\ \citenamefont {Steinberg}}]{Dvir2018}%
  \BibitemOpen
  \bibfield  {author} {\bibinfo {author} {\bibfnamefont {T.}~\bibnamefont {Dvir}}, \bibinfo {author} {\bibfnamefont {F.}~\bibnamefont {Massee}}, \bibinfo {author} {\bibfnamefont {L.}~\bibnamefont {Attias}}, \bibinfo {author} {\bibfnamefont {M.}~\bibnamefont {Khodas}}, \bibinfo {author} {\bibfnamefont {M.}~\bibnamefont {Aprili}}, \bibinfo {author} {\bibfnamefont {C.~H.~L.}\ \bibnamefont {Quay}},\ and\ \bibinfo {author} {\bibfnamefont {H.}~\bibnamefont {Steinberg}},\ }\bibfield  {title} {\bibinfo {title} {Spectroscopy of bulk and few-layer superconducting nbse2 with van der waals tunnel junctions},\ }\href {https://doi.org/10.1038/s41467-018-03000-w} {\bibfield  {journal} {\bibinfo  {journal} {Nature Communications}\ }\textbf {\bibinfo {volume} {9}},\ \bibinfo {pages} {598} (\bibinfo {year} {2018})}\BibitemShut {NoStop}%
\bibitem [{\citenamefont {Sohn}\ \emph {et~al.}(2018)\citenamefont {Sohn}, \citenamefont {Xi}, \citenamefont {He}, \citenamefont {Jiang}, \citenamefont {Wang}, \citenamefont {Kang}, \citenamefont {Park}, \citenamefont {Berger}, \citenamefont {Forr{\'o}}, \citenamefont {Law}, \citenamefont {Shan},\ and\ \citenamefont {Mak}}]{Sohn2018}%
  \BibitemOpen
  \bibfield  {author} {\bibinfo {author} {\bibfnamefont {E.}~\bibnamefont {Sohn}}, \bibinfo {author} {\bibfnamefont {X.}~\bibnamefont {Xi}}, \bibinfo {author} {\bibfnamefont {W.-Y.}\ \bibnamefont {He}}, \bibinfo {author} {\bibfnamefont {S.}~\bibnamefont {Jiang}}, \bibinfo {author} {\bibfnamefont {Z.}~\bibnamefont {Wang}}, \bibinfo {author} {\bibfnamefont {K.}~\bibnamefont {Kang}}, \bibinfo {author} {\bibfnamefont {J.-H.}\ \bibnamefont {Park}}, \bibinfo {author} {\bibfnamefont {H.}~\bibnamefont {Berger}}, \bibinfo {author} {\bibfnamefont {L.}~\bibnamefont {Forr{\'o}}}, \bibinfo {author} {\bibfnamefont {K.~T.}\ \bibnamefont {Law}}, \bibinfo {author} {\bibfnamefont {J.}~\bibnamefont {Shan}},\ and\ \bibinfo {author} {\bibfnamefont {K.~F.}\ \bibnamefont {Mak}},\ }\bibfield  {title} {\bibinfo {title} {An unusual continuous paramagnetic-limited superconducting phase transition in 2d nbse2},\ }\href {https://doi.org/10.1038/s41563-018-0061-1} {\bibfield  {journal} {\bibinfo  {journal} {Nature Materials}\ }\textbf
  {\bibinfo {volume} {17}},\ \bibinfo {pages} {504} (\bibinfo {year} {2018})}\BibitemShut {NoStop}%
\bibitem [{\citenamefont {Xi}\ \emph {et~al.}(2016{\natexlab{b}})\citenamefont {Xi}, \citenamefont {Berger}, \citenamefont {Forr\'o}, \citenamefont {Shan},\ and\ \citenamefont {Mak}}]{Xi2016}%
  \BibitemOpen
  \bibfield  {author} {\bibinfo {author} {\bibfnamefont {X.}~\bibnamefont {Xi}}, \bibinfo {author} {\bibfnamefont {H.}~\bibnamefont {Berger}}, \bibinfo {author} {\bibfnamefont {L.}~\bibnamefont {Forr\'o}}, \bibinfo {author} {\bibfnamefont {J.}~\bibnamefont {Shan}},\ and\ \bibinfo {author} {\bibfnamefont {K.~F.}\ \bibnamefont {Mak}},\ }\bibfield  {title} {\bibinfo {title} {Gate tuning of electronic phase transitions in two-dimensional ${\mathrm{nbse}}_{2}$},\ }\href {https://doi.org/10.1103/PhysRevLett.117.106801} {\bibfield  {journal} {\bibinfo  {journal} {Phys. Rev. Lett.}\ }\textbf {\bibinfo {volume} {117}},\ \bibinfo {pages} {106801} (\bibinfo {year} {2016}{\natexlab{b}})}\BibitemShut {NoStop}%
\bibitem [{\citenamefont {Deng}\ \emph {et~al.}(2018)\citenamefont {Deng}, \citenamefont {Yu}, \citenamefont {Song}, \citenamefont {Zhang}, \citenamefont {Wang}, \citenamefont {Sun}, \citenamefont {Yi}, \citenamefont {Wu}, \citenamefont {Wu}, \citenamefont {Zhu}, \citenamefont {Wang}, \citenamefont {Chen},\ and\ \citenamefont {Zhang}}]{Deng2018}%
  \BibitemOpen
  \bibfield  {author} {\bibinfo {author} {\bibfnamefont {Y.}~\bibnamefont {Deng}}, \bibinfo {author} {\bibfnamefont {Y.}~\bibnamefont {Yu}}, \bibinfo {author} {\bibfnamefont {Y.}~\bibnamefont {Song}}, \bibinfo {author} {\bibfnamefont {J.}~\bibnamefont {Zhang}}, \bibinfo {author} {\bibfnamefont {N.~Z.}\ \bibnamefont {Wang}}, \bibinfo {author} {\bibfnamefont {Z.}~\bibnamefont {Sun}}, \bibinfo {author} {\bibfnamefont {Y.}~\bibnamefont {Yi}}, \bibinfo {author} {\bibfnamefont {Y.~Z.}\ \bibnamefont {Wu}}, \bibinfo {author} {\bibfnamefont {S.}~\bibnamefont {Wu}}, \bibinfo {author} {\bibfnamefont {J.}~\bibnamefont {Zhu}}, \bibinfo {author} {\bibfnamefont {J.}~\bibnamefont {Wang}}, \bibinfo {author} {\bibfnamefont {X.~H.}\ \bibnamefont {Chen}},\ and\ \bibinfo {author} {\bibfnamefont {Y.}~\bibnamefont {Zhang}},\ }\bibfield  {title} {\bibinfo {title} {Gate-tunable room-temperature ferromagnetism in two-dimensional fe3gete2},\ }\href {https://doi.org/10.1038/s41586-018-0626-9} {\bibfield  {journal} {\bibinfo  {journal}
  {Nature}\ }\textbf {\bibinfo {volume} {563}},\ \bibinfo {pages} {94} (\bibinfo {year} {2018})}\BibitemShut {NoStop}%
\bibitem [{\citenamefont {Matsuoka}\ \emph {et~al.}(2023)\citenamefont {Matsuoka}, \citenamefont {Kajihara}, \citenamefont {Wang}, \citenamefont {Iwasa},\ and\ \citenamefont {Nakano}}]{Matsuoka2023}%
  \BibitemOpen
  \bibfield  {author} {\bibinfo {author} {\bibfnamefont {H.}~\bibnamefont {Matsuoka}}, \bibinfo {author} {\bibfnamefont {S.}~\bibnamefont {Kajihara}}, \bibinfo {author} {\bibfnamefont {Y.}~\bibnamefont {Wang}}, \bibinfo {author} {\bibfnamefont {Y.}~\bibnamefont {Iwasa}},\ and\ \bibinfo {author} {\bibfnamefont {M.}~\bibnamefont {Nakano}},\ }\href@noop {} {\bibinfo {title} {Gate-tunable ferromagnetism in a van der waals magnetic semimetal}} (\bibinfo {year} {2023}),\ \Eprint {https://arxiv.org/abs/2304.11890} {arXiv:2304.11890 [cond-mat.mtrl-sci]} \BibitemShut {NoStop}%
\end{thebibliography}%

\end{document}